\documentclass{aa}  

\usepackage{graphicx}
\usepackage{natbib}
\usepackage{txfonts}

\begin{document}
\title{First direct detection of a Keplerian rotating disk around the Be star $\alpha$ Arae using the VLTI/AMBER instrument.}

\titlerunning{First direct detection of a Keplerian disk around  $\alpha$ Arae}

   \author{A.~Meilland  \inst{1}
   	 \and
            Ph.~Stee\inst{1}
            \and
            M.~Vannier\inst{3}
            \and
            F. ~Millour \inst{2}
            \and
          A.~Domiciano~de~Souza \inst{1,} \inst{2}
          \and
          F.~Malbet \inst{6}
          C.~Martayan \inst{4}
          \and
          F.~Paresce \inst{7}
          \and
          R.~Petrov \inst{2}
          \and
          A.~Richichi \inst{5}
          \and
          A.~Spang \inst{1} 
          }

   \offprints{anthony.meilland@obs-azur.fr}

\institute{Observatoire de la C\^{o}te d'Azur-CNRS-UMR 6203, D\'epartement GEMINI,  Avenue Copernic, Grasse, France
 \and
Laboratoire Universitaire d'Astrophysique de Nice, Parc Valrose, 06108 Nice Cedex 02, France
 \and
European Southern Observatory, Cerro Paranal, Chile
 \and
GEPI-Observatoire de Paris-Meudon, 5, Place Jules Janssen, 92195 Meudon Cedex, France
 \and
European Southern Observatory Karl-Schwarzschild-Strasse 2 D-85748 Garching bei M\"unchen, Germany
 \and
Laboratoire d'Astrophysique de l'Observatoire de Grenoble, 414, Rue de la Piscine,
Domaine Universitaire, BP 53, 38041 Grenoble Cedex 09, France
\and
I.N.A.F., Via del Parco Mellini 84, Rome, Italy
}

   \date{Received; accepted }

   \abstract{}{We aim to study the geometry and kinematics of the disk around the Be star $\alpha$ Arae as a function of wavelength, especially across the Br$\gamma$ emission line. The main purpose of this paper is to answer the question about the nature of the disk rotation around Be stars.}{We use the VLTI/AMBER instrument operating in the K band which provides a gain by a factor 5 in spatial resolution compared to previous VLTI/MIDI observations. Moreover, it is possible to combine the high angular resolution provided with the (medium) spectral resolution of AMBER to study the kinematics of the inner part of the disk and to infer its rotation law. }{We obtain for the first time the direct evidence that the disk is in keplerian rotation, answering a question that occurs since the discovery of the first Be star $\gamma$ Cas by father Secchi in 1866. We also present the global geometry of the disk showing that it is compatible with a thin disk + polar enhanced winds modeled with the SIMECA code. We found that the disk around $\alpha$ Arae is compatible with a dense equatorial  matter confined in the central region whereas a polar wind is contributing along the rotational axis of the central star. Between these two regions the density must be low enough to reproduce the large visibility modulus (small extension) obtained for two of the four VLTI baselines. Moreover, we obtain that $\alpha$ Arae is rotating very close to its critical rotation. This scenario is also compatible with the previous MIDI measurements.}{}


   \maketitle
%

\section{Introduction}
The star $\alpha$ Arae (HD\,158\,427, HR\,6510, B3\,Ve), one of the closest (d=74 pc, Hipparcos, Perryman et al. \cite{perryman}) Be stars, was observed with the VLTI/MIDI 
instrument at 10 $\mu$m in June 2003 and its circumstellar environment was unresolved even 
with the 102m baseline (Chesneau et al. \cite{chesneau}, hereafter paper I). $\alpha$ Arae was a natural 
choice as first target due to its proximity but also its large mid-IR flux and its high infrared
excess among other Be stars, e.g. E(V-L)$\sim$1.8 and E(V-12$\mu$m)$\sim$2.23. These first IR interferometric measurements indicated that the size of the circumstellar environment
was smaller than predicted by
Stee \cite{Stee4} for the K band. The fact that $\alpha$~Arae remain
unresolved, but at the same time had strong Balmer emission, have put very strong
constraints on the parameters of its circumstellar disk. Independently of the
model, they have obtained an upper limit of the envelope size in the N band
of $\phi_{max}$= 4 mas, i.e. 14 R$_{\star}$ if the star is at 74 pc according
to Hipparcos parallax or 20 R$_{\star}$ if the star is at 105 pc as suggested
by the model presented in paper I. \\

They finally propose a scenario where the circumstellar environment 
remains unresolved due to an outer truncation of the disc by an unseen companion.
Nevertheless, this companion would be too small and too
far away to have any influence on the Be phenomenon itself.\\

In order to study the inner part of this circumstellar truncated disk we have taken
advantage of the higher spatial resolution by observing at 
2 $\mu$m with the VLTI/AMBER instrument in February 2005. It  provides a gain by 
a factor 5 in spatial resolution compared to VLTI/MIDI observations. We present in this 
paper these measurements showing, for the first time, a fully resolved circumstellar 
envelope in the Br$\gamma$ emission line and a clear signature of a Keplerian rotating disk
around $\alpha$ Arae. We also discuss the challenging question on the nature of the
geometry of the Be disks and particularly their opening angle since it is
still an active debate.\\

Following the Wind Compressed Disk model
(WCD) by Bjorkman \& Cassinelli \cite{Bjorkman93}, most authors
have considered geometrically thin disks (half opening angle of
2-5 degrees) even if Owocki et al. (1996) have found that the
equatorial wind compression effects are suppressed in any
radiatively driven wind models for which the driving forces
include a significant part from optically thick lines. Moreover,
they found that gravity darkening effects can lead to a reduced
mass loss, and thus a lower density in the equatorial regions. A
wind compression effect is, however, not required to produce small
opening angle of the disk. The investigation of accretion disks
has shown that discs in hydrodynamical equilibrium and Keplerian
rotation will not have much larger opening angles, since their
scale height is governed by the vertical gas pressure only. For a
disk to be thicker, either additional mechanisms have to be
assumed, or it might not be in equilibrium at the radii in
question (Bjorkman \& Carciofi 2004).\\

On the other side, Stee et al. (1995; 1998) claimed that Be disks
must be more ellipsoidal in order to reproduce the strong IR
excess observed and interpret the possibility for a Be star to
change its spectral type from Be to B, and more rarely Be to
Be-Shell type, i.e. where the disk is dense enough to produce a
strong "shell" absorption.\\

In the following we adopt, as a starting point,
the same parameters for the modeling of $\alpha$ Arae, e.g.
the central star and its circumstellar envelope used in paper I and
summarized in Table 1.  Following the polarization measurements 
Pl$\approx$0.6\% and Position Angle (PA) of 172$^\circ$ by 
McLean\&Clarke \citealp{mclean} and Yudin
\citealp{yudin2} the disk major-axis orientation is expected to be
at about PA$\approx82^\circ$ (Wood et al.  \citealp{wood96a},
\citealp{wood96b}, Quirrenbach et al. \citealp{quirrenbach2}).
Assuming a stellar radius of 4.8~R$_{\sun}$ and an effective
temperature $T_{\rm eff}=18\,000$\,K, the photospheric angular
diameter is estimated to be 0.7~mas (Cohen et al. \citealp{cohen},
Chauville et al. \citealp{chauville}). For the distance of 74~pc and a baseline 
of 60m at 2~$\mu$m, Stee \cite{Stee4} predicts the visibility of $\alpha$~Arae 
to be lower than 0.2, i.e. fully resolved.\\

The paper is organized as follows. In Section~\ref{secobs} and \ref{datared} we present
the interferometric AMBER observations and the data reduction.
In Section~\ref{toymodels} we try to obtain a first estimate of $\alpha$ Arae's envelope geometry 
using very simple "toy" models. Section~\ref{secsimeca} describes briefly the SIMECA code.
In Section \ref{bestmodel} we present the best model we obtain with SIMECA that fits both the 
Br$\gamma$ line and the visibility modulus and phase as a function of wavelength which allows 
us to infer the disk kinematics and its rotational velocity. Finally, Section \ref{conclusion} draws the
conclusions from these first spectrally resolved interferometric measurements of a 
Be star at 2 $\mu$m.

\vspace{0.3cm}
{
\begin{table}
{\centering \begin{tabular}{cc} \hline
parameter/result    & value \\
\hline
Spectral type& B3Ve\\
$T_{\rm eff}$& 18\,000\,K\\
Mass& 9.6 M\( _{\sun } \)\\
Radius& 4.8 R\( _{\sun } \)\\
Luminosity& 5.8 10\( ^{3} \)L\( _{\sun } \)\\
Inclination angle i & 45$\degr$\\
Photospheric density ($\rho_{phot}$)&1.2 10\( ^{-12} \)g cm\( ^{-3} \)\\
Photospheric expansion velocity& 0.07 km s\( ^{-1} \) \\
Equatorial rotation velocity & 300 km s\( ^{-1} \)  \\
Equatorial terminal velocity & 170 km s\( ^{-1} \) \\
Polar terminal velocity & 2000 km s\( ^{-1} \) \\
Polar mass flux & 1.7 10\( ^{-9} \)M\( _{\sun } \) year\( ^{-1} \) sr\( ^{-1} \) \\
m1 & 0.3 \\
m2 & 0.45 \\
C1 & 30\\
Mass of the disk & 2.3 10\( ^{-10} \)M\( _{\sun } \) \\
Mass loss & 6.0 10\( ^{-7} \)M\( _{\sun } \) year\( ^{-1} \)\\
\hline
\end{tabular}\par}

\caption{Model parameters for the $\alpha$ Arae central star and its circumstellar environment
obtained from paper I}
\label{midi_model}
\end{table}
\par}
\vspace{0.3cm}

\section{VLTI/AMBER observations}
\label{secobs}
Our observations of $\alpha$ Arae were acquired during a Science Demonstration Time (SDT) run, on the nights of 23 and 24 February  2005, using the AMBER 
instrument on the VLTI in medium-resolution mode (R=1500). On the night of 24 February, the observations were made with two UT telescopes, i.e. one interferometric
baseline only,  and consist of six exposure files, each containing 500 frames of 100\,ms. On the following night, three telescopes were used and
three-baselines data were taken in a serie of three exposure files and another serie of two, each containing 500 frames of 70\,ms. Without a fringe tracker,
the integration time per frame must be short enough to minimize the smearing of the fringe visibility due to the beams jitter, while still having enough
photons over the elementary exposure. Its specific value is chosen depending on the atmospheric conditions. Immediately after observing $\alpha$ Arae, a
nearby calibrator object was observed.

\begin{table*}[t]
{\centering
\begin{tabular}{ccccccc} \hline
Date       & Time${\rm obs}$ & Baselines & Length & P.A. & Calibrator & $\phi_{\rm cal.}$\\
\hline
   & (UTC)        &    & (m) & ($^o$) & &    (mas)    \\   24/12/2005   & 08:46 &   UT2-4 (B$_0$)  & 80.9 & 39 & HD124454 &  $1.52\pm 0.02$\\
   &  &  UT2-3 (B$_1$)& 46.4 & 19 &  & \\
   &  &  UT3-4 (B$_2$)& 52.5 & 81 &  & \\
   25/12/2005   & 09:40 &  UT2-4 (B$_3$)& 84.6 & 52 & HD124454  &   $1.52\pm 0.02$\\
\hline
\end{tabular}
\par}
\caption{Observation log, with the projected baseline lengths and angles. The diameter for the calibrator used is taken from the CHARM catalogue (Richichi \& Percheron \citealp{Richichi02}).}
\label{Table_extension}
\end{table*}

\section{Reduction of the interferometric Data}
\label{datared}
The data has been reduced using the ``ammyorick'' package developped by the AMBER
consortium\footnote{see: http: $//$www$-$laog.obs.ujf$-$grenoble.fr$/$heberges/amber\\/article.php3?id$\_$article=81}. The principles of the AMBER data reduction have been described by Millour et al (2005).  In addition to the tools furnished by the default package, some specific treatment have been added for reaching an optimal precision on the interferometric observables. The various steps are the following. For each of the individual exposure frames, the
complex visibilities are extracted from the various interferometric channels and calibrated using the photometric channels and some internal calibration files. 
The piston $p$  between the beams is first estimated from the slope of the fringes with wavelength, and a correction phaser is applied on each frame.
From this, the color-differential phase and visibility are calculated, for a given spectral channel $\lambda$, with respect to a set of reference channel(s).
We chose the reference channel to consist in the whole spectral bandwidth except the considered channel $\lambda$.  Due to the jitter of the beams and
to the subsequent variations of output flux after the optical fibers of AMBER, not all the frames contain good quality fringes.  A selection and weighing
of the best frames above a given threshold is made based on a fringe quality criterion. This yields our color-differential estimators of phase and visibility
 with an optimal precision.  For both the science source and its calibrator, the statistical deviation of the differential phases over the frames serie
is about 0.01\,rad per spectral channel. The deviation of the normalized differential visibility is about $0.5\%$. These numbers are similar for all the
baselines.

The scaling of  the differential visibilities to ``absolute'' visibilities is done through a best adjustment of the ``science'' data to the one measured on
the calibrator star, whose theoretical visibility is known. For this, the histogram of the squared visibilities from all the science frames (with enough
flux above the detector noise) is fitted, by a scaling factor, to the histogram of the calibrator star. With the data we have, this fit has been done globally,
by integrating the spectral channels together for each frame. Therefore, some possible bias between the spectral channels would not be suppressed by
the calibration process.

In the present case, the observations of the nominal calibrator for $\alpha$ Arae were of poor quality, and another calibrator, at some distance on the sky
($\approx$ 45$\degr$), had to be used. This does not seem to affect the calibration, though. The scaling fit between the histograms of the science and 
calibrator objects results in an error of about $5\%$ on the visibilities, on every baseline.

\section{Study of the envelope morphology using simple models}
\label{toymodels}
In this section we present the new AMBER data with the previous MIDI data already presented in Paper I in order to obtain a first
estimate of $\alpha$ Arae's envelope geometry using very simple "toy" models.

\begin{figure}
	\begin{center}
	\includegraphics[height=7.0cm]{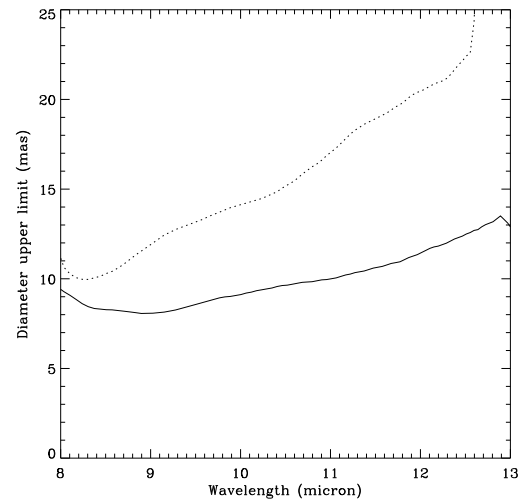}
      \caption{$\alpha$ Arae (unresolved star + uniform disc) model upper limit diameters (in mas), as a function of wavelength, from the 2005 MIDI data described in paper I for the June, 16 (plain line) and June, 17 nights (dotted line). Note that the large values obtained for larger wavelengths are essentially due to a poor calibration and large error bars  rather than a true physical effect.}
   \end{center}
  \label{midi_results}
\end{figure}

\label{envmorph}
\subsection{Envelope extension}
\subsubsection{Extension in the continuum at 2.1 $\mu$m}
Assuming that the visibility we measure in the continuum, $V_c$, is only due to the central star and its circumstellar disk we can write: 

\begin{equation}
V_c= \frac {V_{ec} F_{ec} + V_{\star c} F_{\star c}} {F_{tot}}
\label{Eq.1}
\end{equation}
 where V$_{ec}$ and F$_{ec}$ are respectively the envelope visibility and flux in the continuum , V$_{\star c}$ and F$_{\star c}$ the star visibility and flux in the continuum and F$_{tot}$ = F$_{ec}$+F$_{\star c}$. Since V$_c$ is the measured visibility, in order to only estimate the envelope visibility, V$_{ec}$, we can rewrite Eq. \ref{Eq.1} :
\begin{equation}
V_{ec}= \frac {V_c F_{tot} - V_{\star c} F_{\star c}} {F_{ec}}
\label{Eq.2}
\end{equation}
The total flux is normalized, i.e. F$_{tot}$=F$_{ec}$+F$_{\star c}$=1. Since the star is almost unresolved 0.5$<$$\phi_{\star}$$<$0.7 mas which corresponds to 0.98$<$ V$_{\star c}$$<$0.99 for the longest baseline at 2.1 $\mu$m, we assume in the following that V$_{\star c}$=1. In order to estimate V$_{ec}$ we still have to determine the star and envelope contributions to the 2.1 $\mu$m flux continuum.  
Since the envelope continuum flux in the visible and in the UV is negligible we have fitted the blue part of the SED using a T$_{eff}$=18000 K black body for the central star in order to deduce the envelope emission at larger wavelength.  At 2.1$\mu$m  we found that the star emission is still $\sim$1.5 larger, i.e. 0.44 magnitude brighter,  than the envelope contribution thus, following F$_{\star c}$=1.5 F$_{ec}$, we obtain
F$_{\star c}$=0.6 and F$_{ec}$=0.4.
Equation \ref{Eq.2} can now be rewritten as:
\begin{equation}
V_{ec}= 2.5V_c -  1.5
\label{Eq.3}
\end{equation}
The continuum visibilities and envelope extensions we obtained following Eq. \ref{Eq.3}, assuming a uniform disk model, are given in table \ref{Table_extension}.

\subsubsection{Extension in the Br$\gamma$ line}
We can define the same equation as Eq. \ref{Eq.2} for the envelope visibility in the Br$\gamma$ line following:
\begin{equation}
V_{er}= \frac {V_r F_r - V_c F_{tot}} {F_{er}}
\label{Eq.4}
\end{equation}
where V$_r$ and F$_r$ are respectively the measured visibility and flux in the Br$\gamma$ line. V$_c$ and F$_{tot}$ are the quantities previously defined and V$_{er}$ and F$_{er}$ are the 
visibility and flux only due to the envelope, i.e. without the stellar contribution. We obtain using the Br$\gamma$ emission line profile F$_{er}$=0.5 and F$_r$=1.5 at the center of the line.
The values of V$_{er}$ we obtain using Eq. \ref{Eq.4} assuming a uniform disk model, are given in Table \ref{Table_extension}.

\noindent Globally, we obtain that, using a simple uniform disk model for each measurement, the Br$\gamma$ emission in the line and the nearby continuum at 2.1 $\mu$m have the same extension about 7-8 mas along the UT2-UT3 (B$_1$)and UT3-UT4 (B$_2$) baselines and about 5 mas for UT2-UT4 (B$_3$) and $\sim$ 4 mas for UT2-UT4 (B$_0$) (which was measured at a slightly different PA).

\subsection{Comparison between MIDI and AMBER extensions}
In order to compare the extensions we obtain with AMBER at 2.1 $\mu$m (see Table  \ref{Table_extension}) with the MIDI data presented in paper I, we have calculated the envelope extension in the continuum
between 7 and 13 $\mu$m following Eq. \ref{Eq.2}, assuming  a uniform disk for each measurement. Unfortunately, these MIDI observations were conducted under unfavorable conditions with thin cirrus passing, and the data were reduced with one of the first versions of the software available.  This led to error bars between 8\% and 18\%, insufficient to measure a significant change of the visibility and thus of the diameter between 7 and 13 $\mu$m. Despite these limitations, these measurements have roughly given an upper limit value of the envelope extension around 10 mas in the N band (see Fig.~1) and are compatible with our AMBER measurements. 

Moreover, the extension of $\alpha$ Arae circumstellar environment seems to be almost independent on the wavelength which puts strong constraints on the density law within the envelope.  The large upper limit values obtained for larger wavelengths are essentially due to a poor calibration and large error bars  rather than a true physical effect. For "classical" Be stars models, where the density is slowly and regularly decreasing as a function of the distance from the central star, we predict an increasing size as a function of wavelength as shown by Stee \cite{Stee3}. A possibility to keep the same angular size between 2 and 13 $\mu$m, may be a  disk truncation by an external physical effect as proposed in paper I. 

Using FEROS spectroscopic data and the fact that their VLTI/MIDI data are showing a nearly unresolved envelope they propose a possible disk truncation by a unseen companion at a radius of about $154\,{\rm R}_\odot$, assuming a circular orbit for the companion with negligible mass. With $R_\star=4.8\,{\rm R}_\odot$, this corresponds to about 32 stellar radii, i.e. 6.4 mas, which is in agreement with their estimate based on the VLTI/MIDI data for a disk truncated at 25~R$_\star$, i.e. 4 mas \ somewhat smaller than the companion orbit. This is also in agreement with what we have obtained with AMBER, i.e. an envelope extension at 2.1 $\mu$m around 7 mas which may appears larger than the 4 mas they obtained. Nevertheless, we must keep in mind that they estimate their extensions using a uniform disk for the star+disk emission whereas we use a unresolved star + a uniform disk only for the circumstellar envelope leading to a larger size. 


Finally, we have verified that our best model described in Section~\ref{bestmodel} was compatible with the MIDI data obtain in 2003 showing a nearly unresolved envelope.

\begin{figure}[h]
	\begin{center}
			\includegraphics[height=7.0cm]{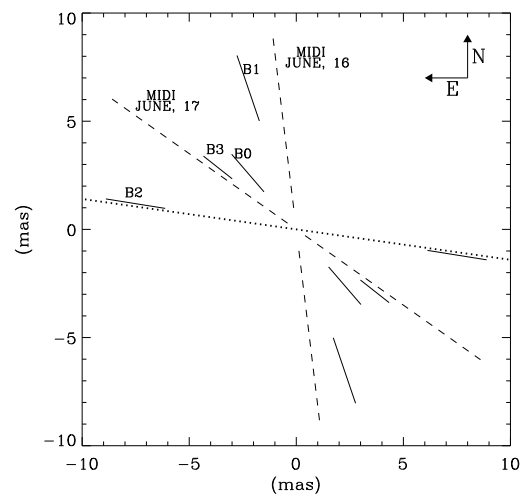}			
			\caption{ $\alpha$ Arae (unresolved star + uniform disc) model diameters (in mas) dependence on the baseline position angle. Full line: AMBER measurements in the continuum at 2.1 $\mu$m. Dashed lines: MIDI measurements. The length of the plot corresponds to the values of the error bars (note the very large error bars for the MIDI data).The dotted line is the direction of the major-axis of the envelope estimated from polarization measurements (PA=172$\degr$) obtained by McLean \& Clarke \cite{mclean} and Yudin et al. \cite{yudin}}		
   \end{center}
 \label{skyplane_extension}
 \end{figure}

{
\begin{table}[h]
{\centering \begin{tabular}{c|cccc} \hline
 Base n$^o$ & 0 & 1 & 2 & 3 \\

 \hline
 Description & UT2-4 & UT2-3 & UT3-4 & UT2-4\\
 Length (m)& 80.9 &  46.4 & 52.5 & 84.6 \\
 P.A. ($^o$) & 39 & 19 & 81 & 52 \\
 \hline
 V$_c$ &0.84&0.80&0.72&0.73 \\
  \hline
 V$_r$ &0.63&0.70&0.59&0.55\\
 \hline
 V$_{ec}$ &0.60&0.50&0.30&0.33\\
 \hline
 V$_{er}$ &0.21&0.50&0.33&0.19\\
 \hline
 $\phi$$_{ec}$ (mas)&3.3 $\pm$1.2&6.8$\pm$1.6&7.6$\pm$1.4&4.5$\pm$0.9\\
 \hline
 $\phi$$_{er}$ (mas)&5.3$\pm$2.0&6.8$\pm$1.4&7.3$\pm$2.1&5.2$\pm$1.2\\
\hline
\end{tabular}\par}
\caption{Visibilities measured in the continuum at 2.1 $\mu$m (V$_c$) and the Br$\gamma$ line (V$_r$) and deduced for the envelope contribution only, using a uniform disk model respectively in the continuum
(V$_{ec}$) and in the  Br$\gamma$ line (V$_{er}$). The corresponding angular diameters obtained for the envelope are given in the continuum $\phi$$_{ec}$ and in the line $\phi$$_{er}$.}
\label{Table_extension}
\end{table}
}

\begin{figure*}
	\begin{center}
			\includegraphics[height=4.15cm]{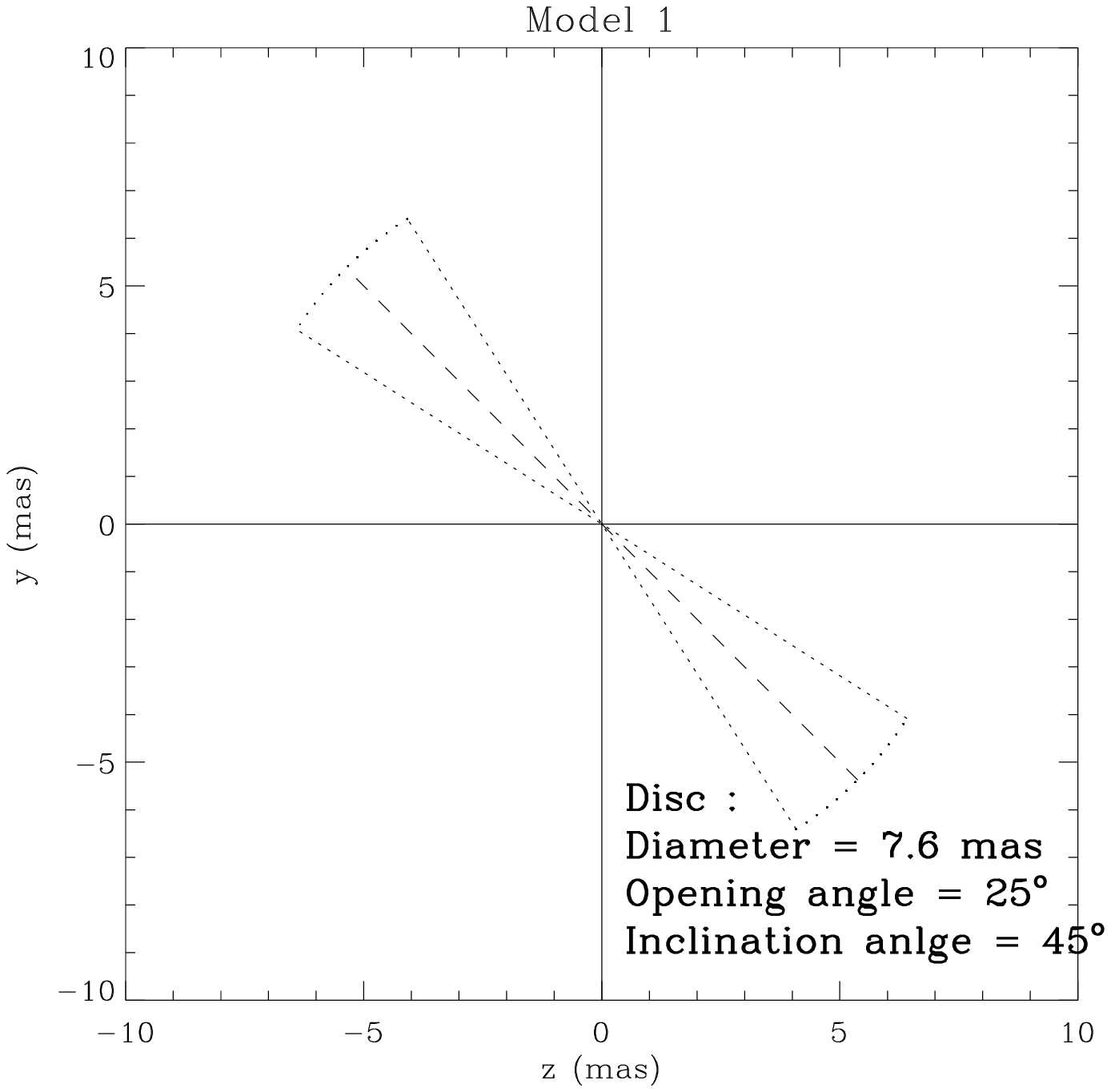}
			\includegraphics[height=4.15cm]{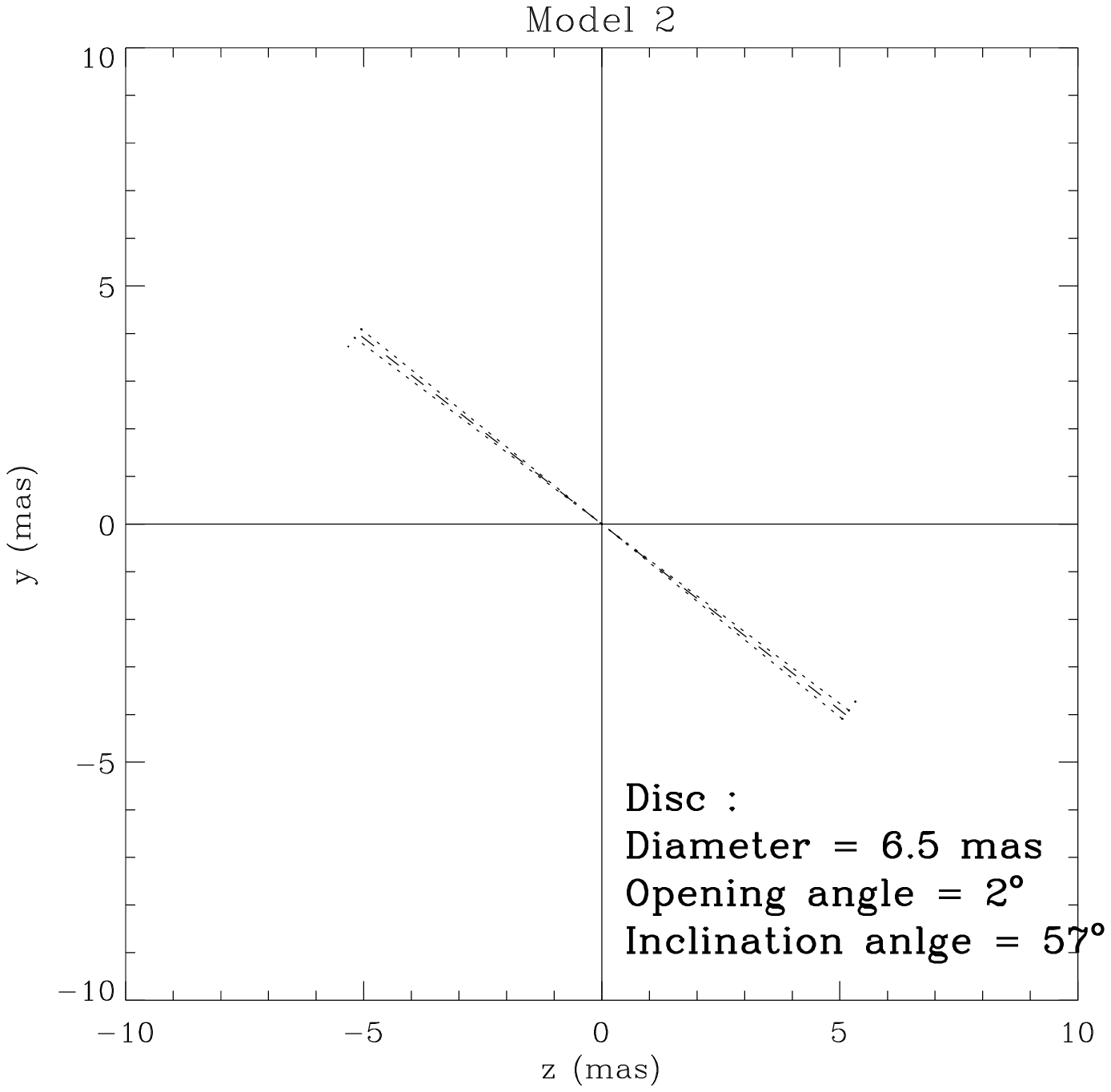}
			\includegraphics[height=4.15cm]{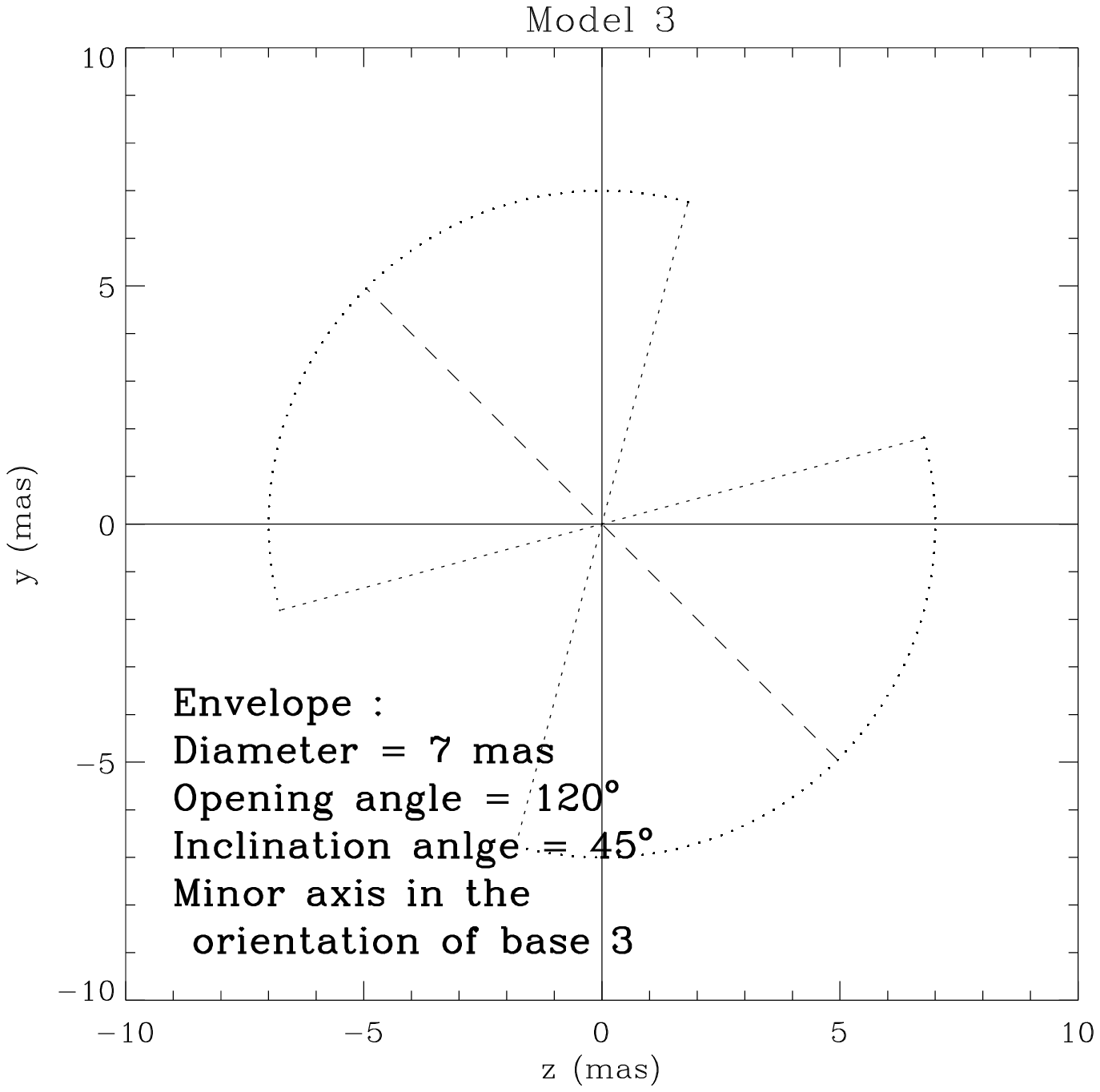}
			\includegraphics[height=4.15cm]{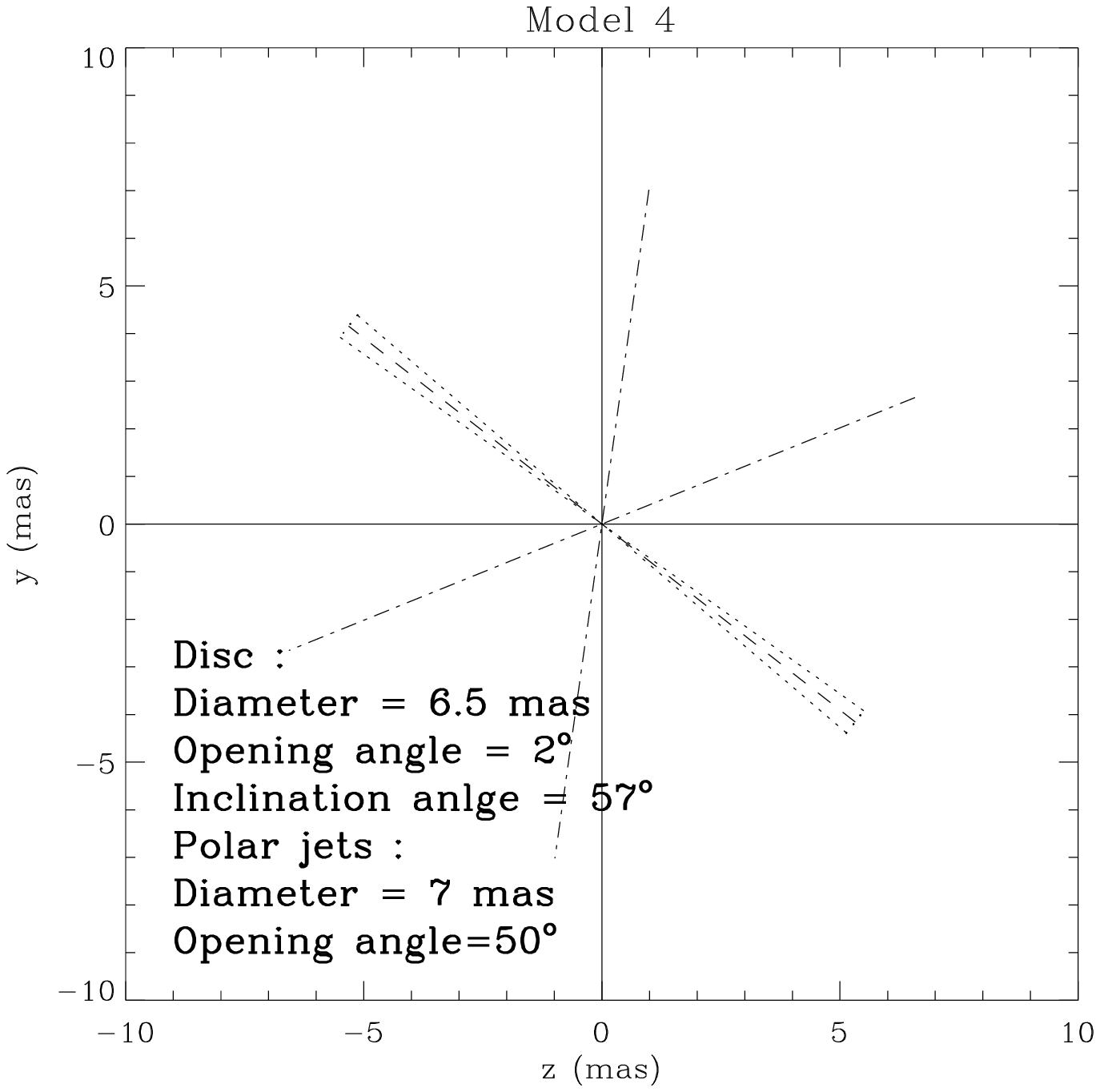}
			\includegraphics[height=4.15cm]{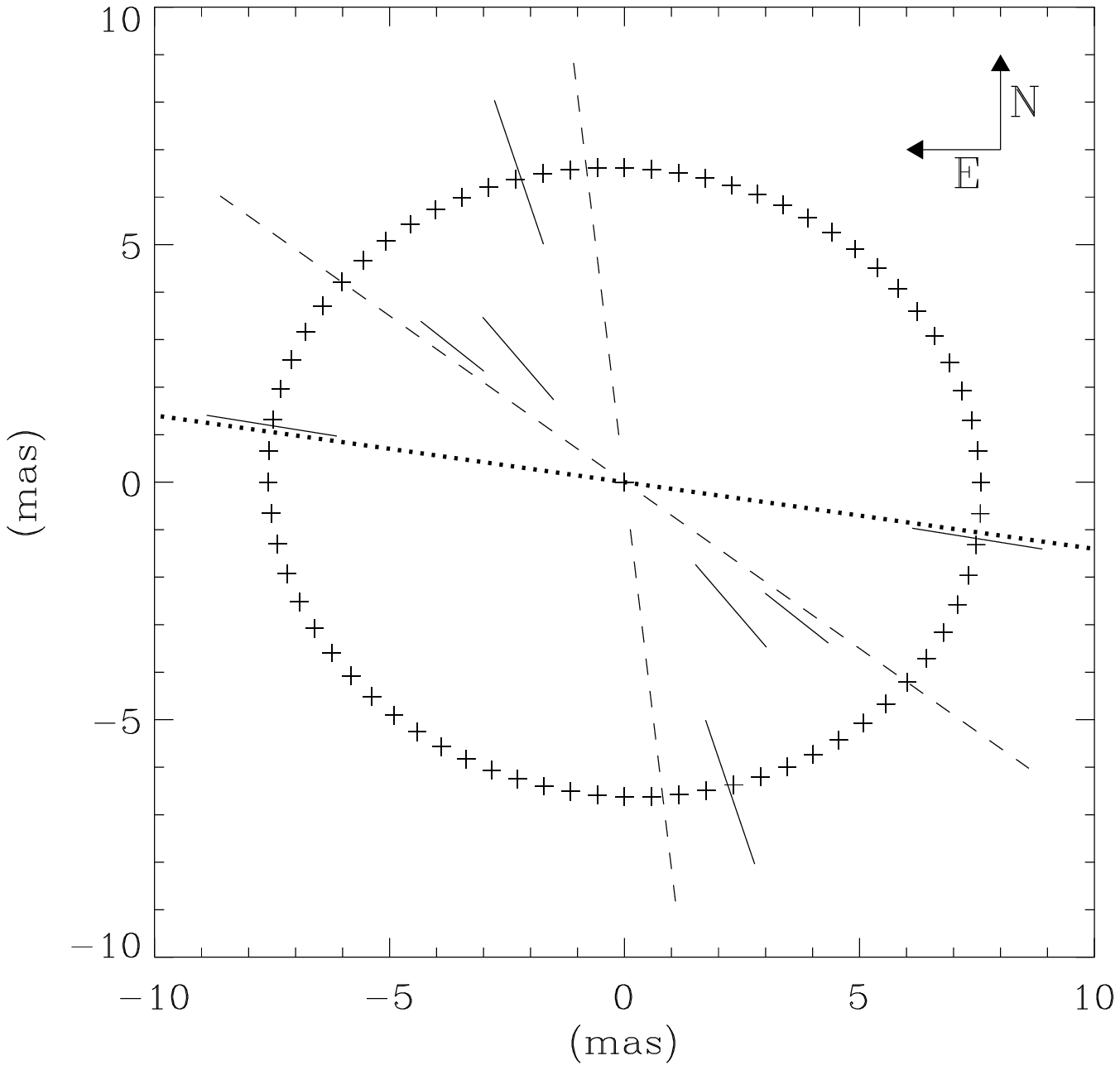}
			\includegraphics[height=4.15cm]{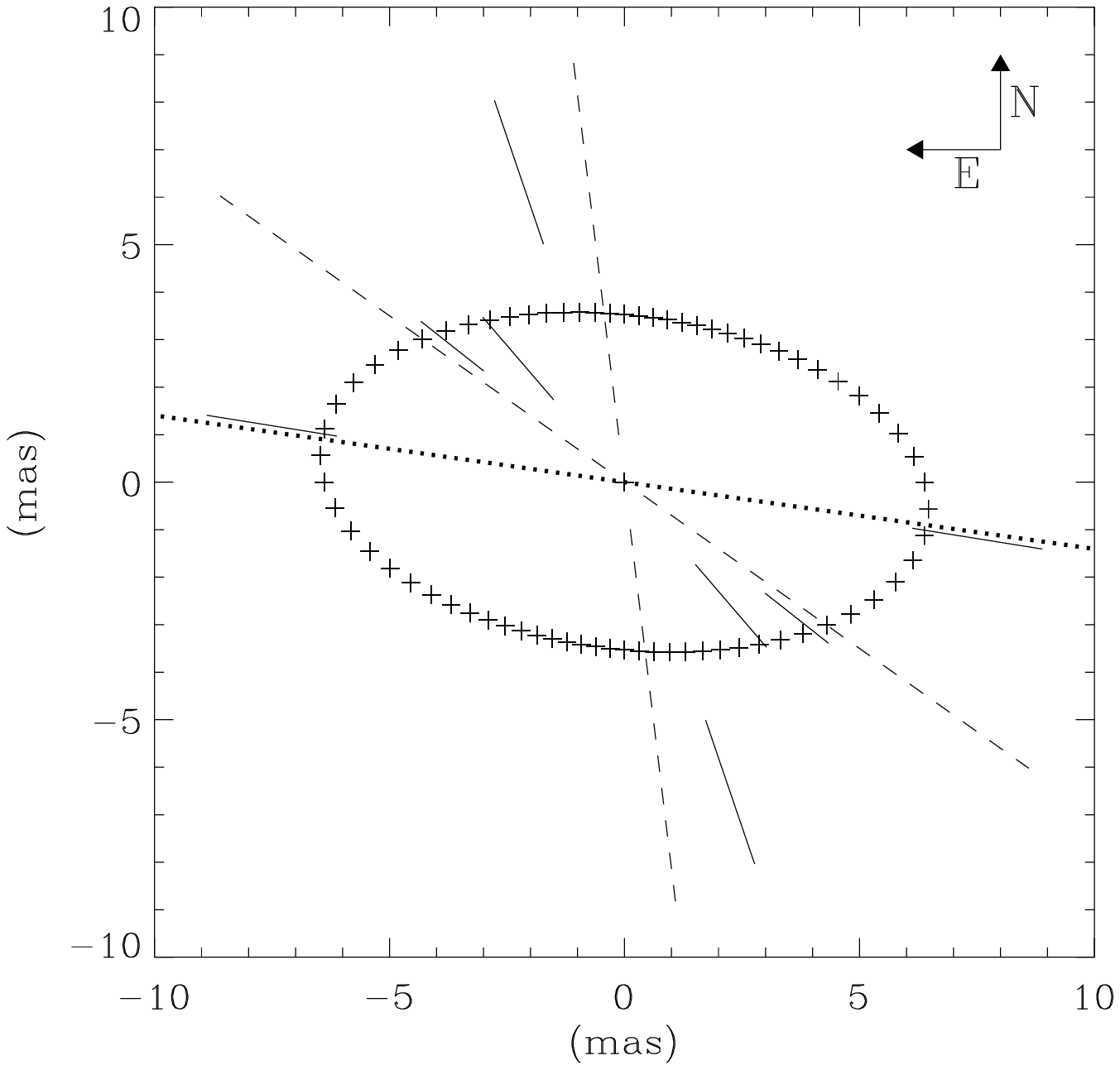}
			\includegraphics[height=4.15cm]{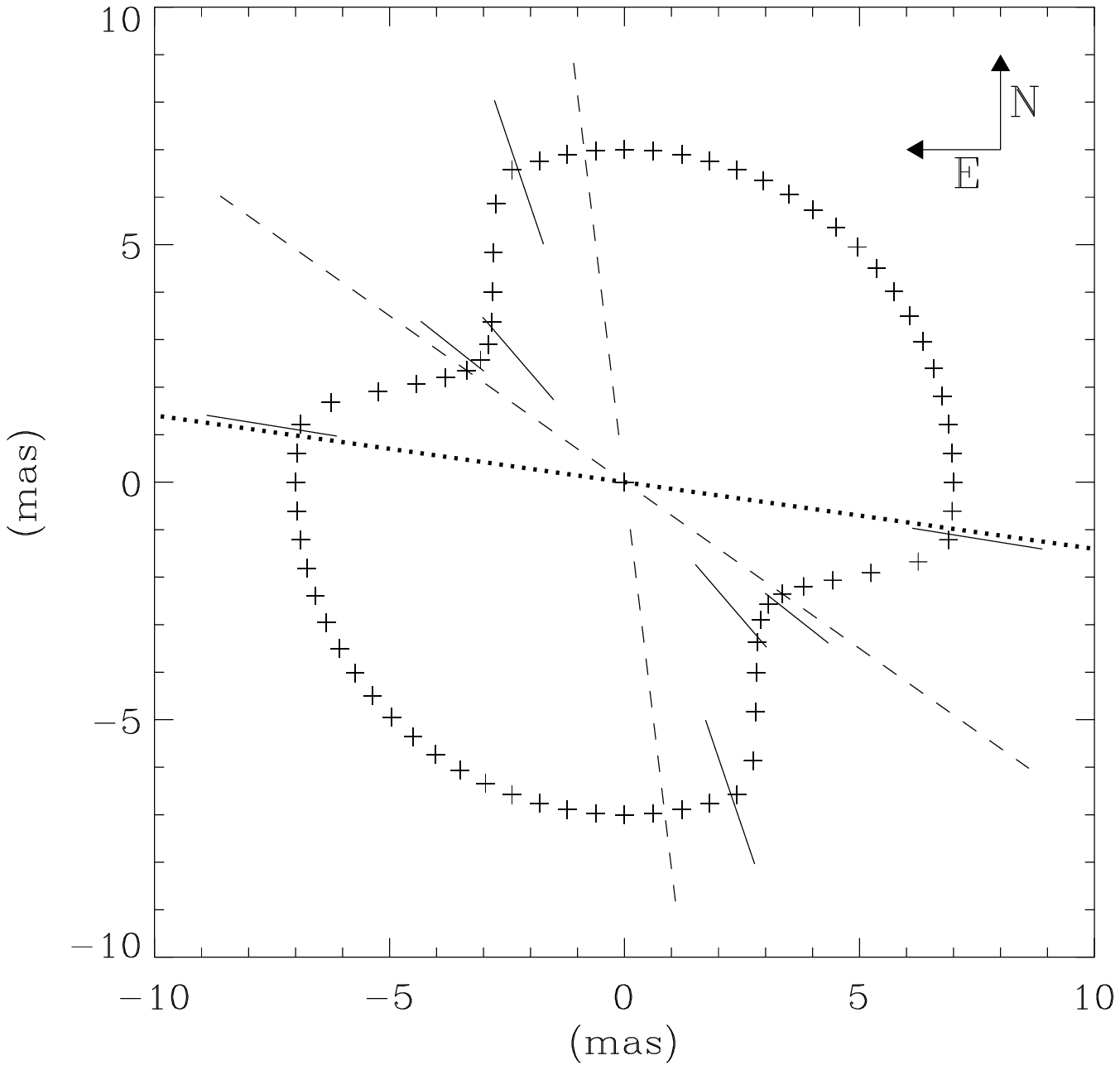}
			\includegraphics[height=4.15cm]{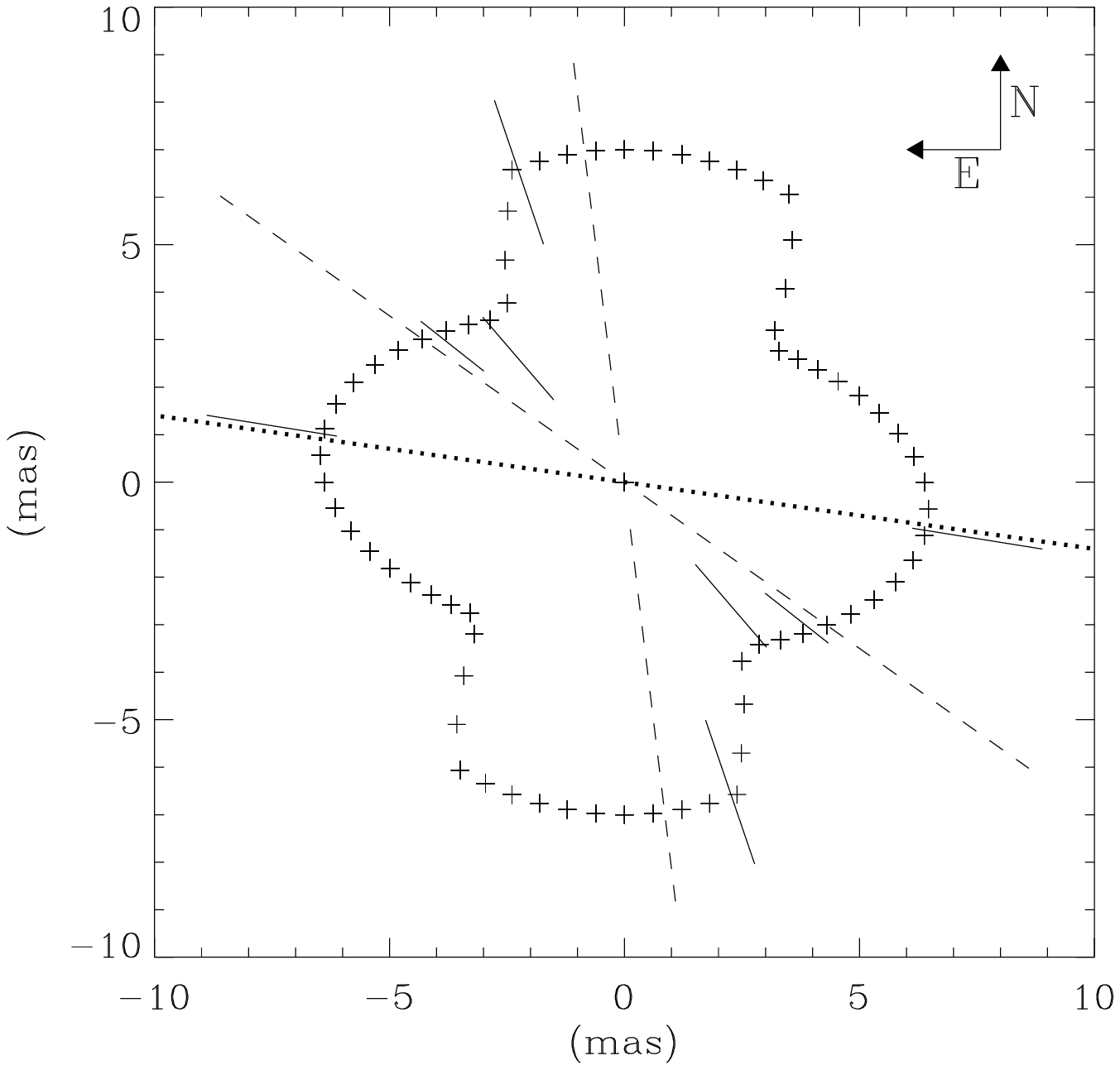}
			\includegraphics[height=4.15cm]{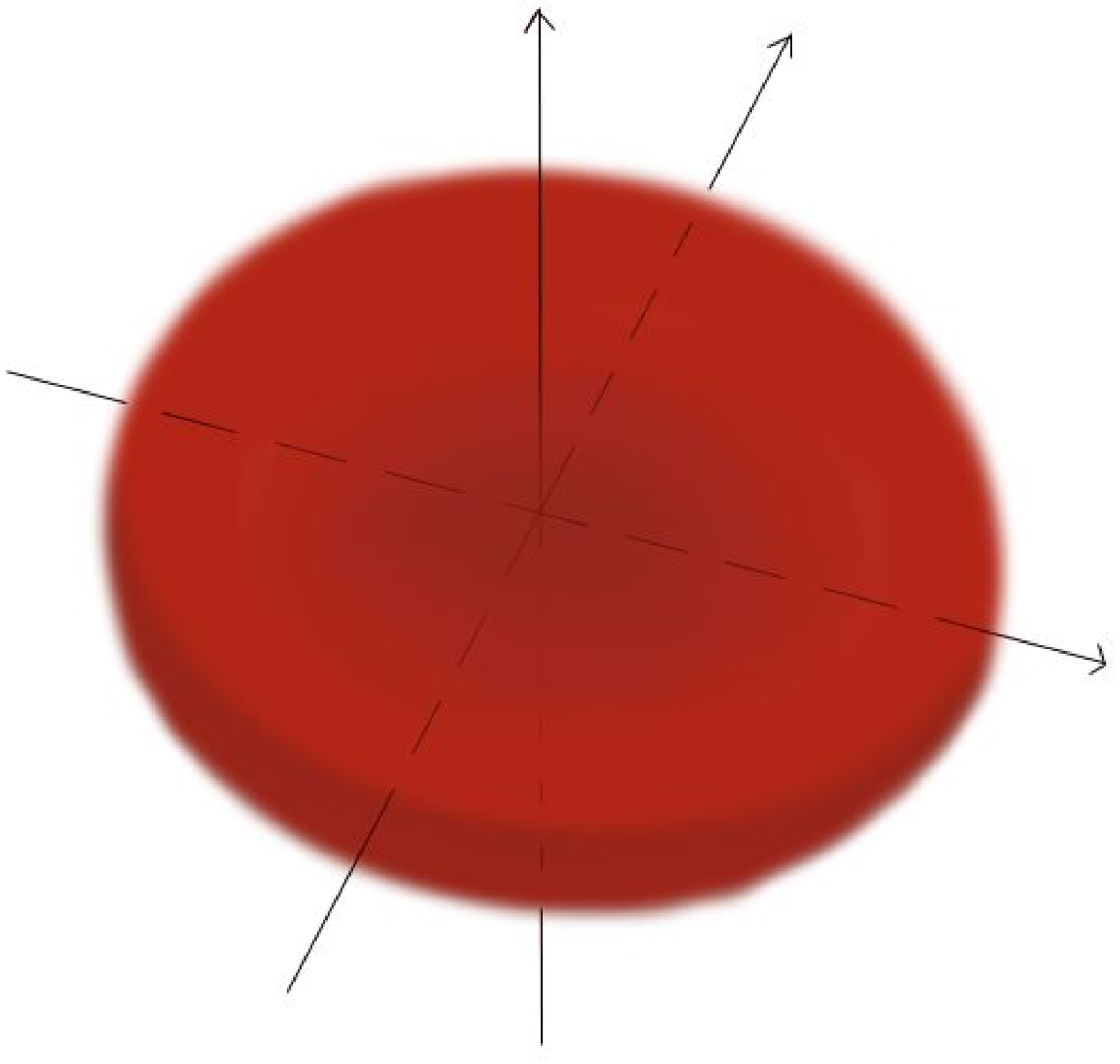}
			\includegraphics[height=4.15cm]{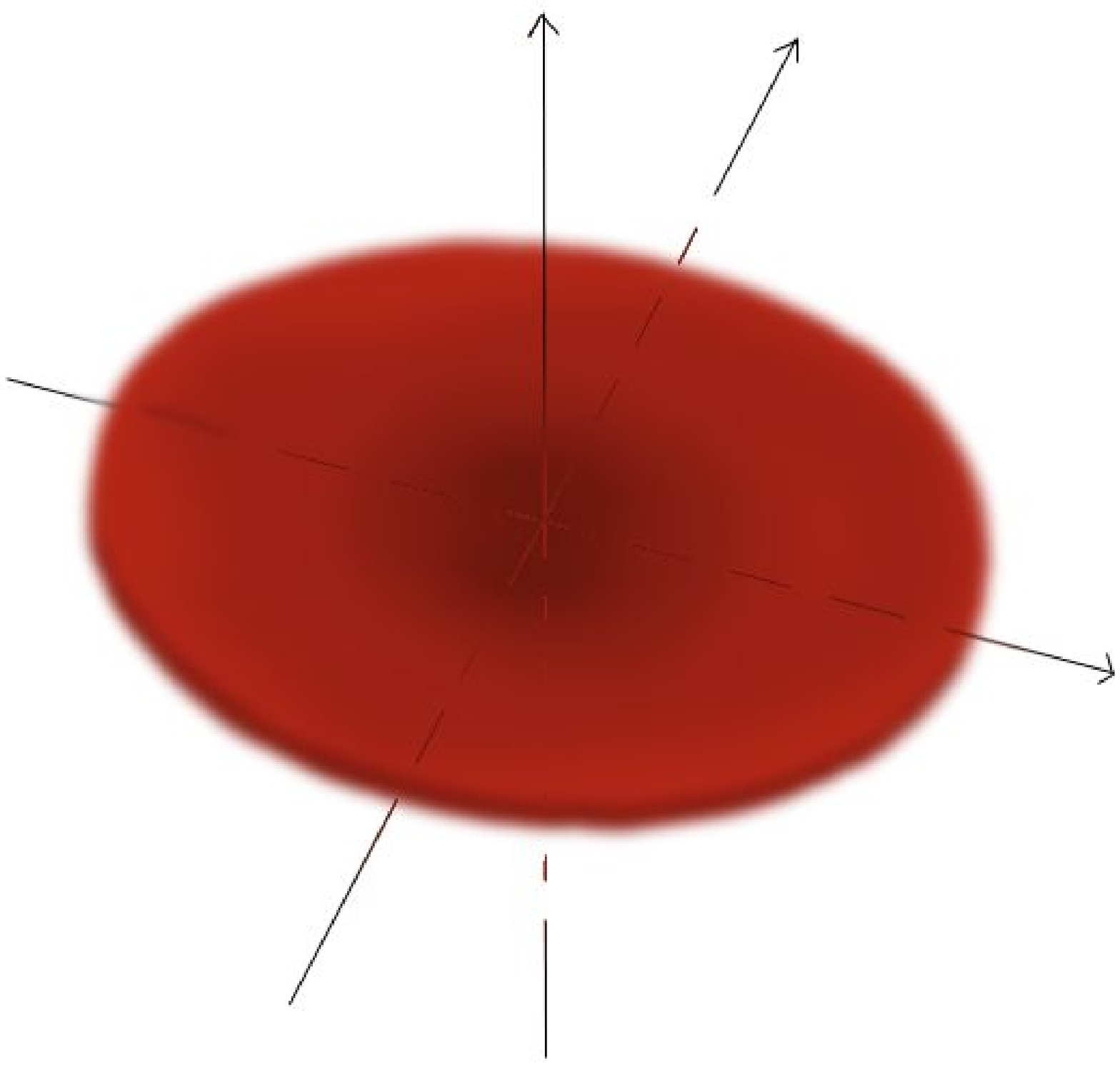}
			\includegraphics[height=4.15cm]{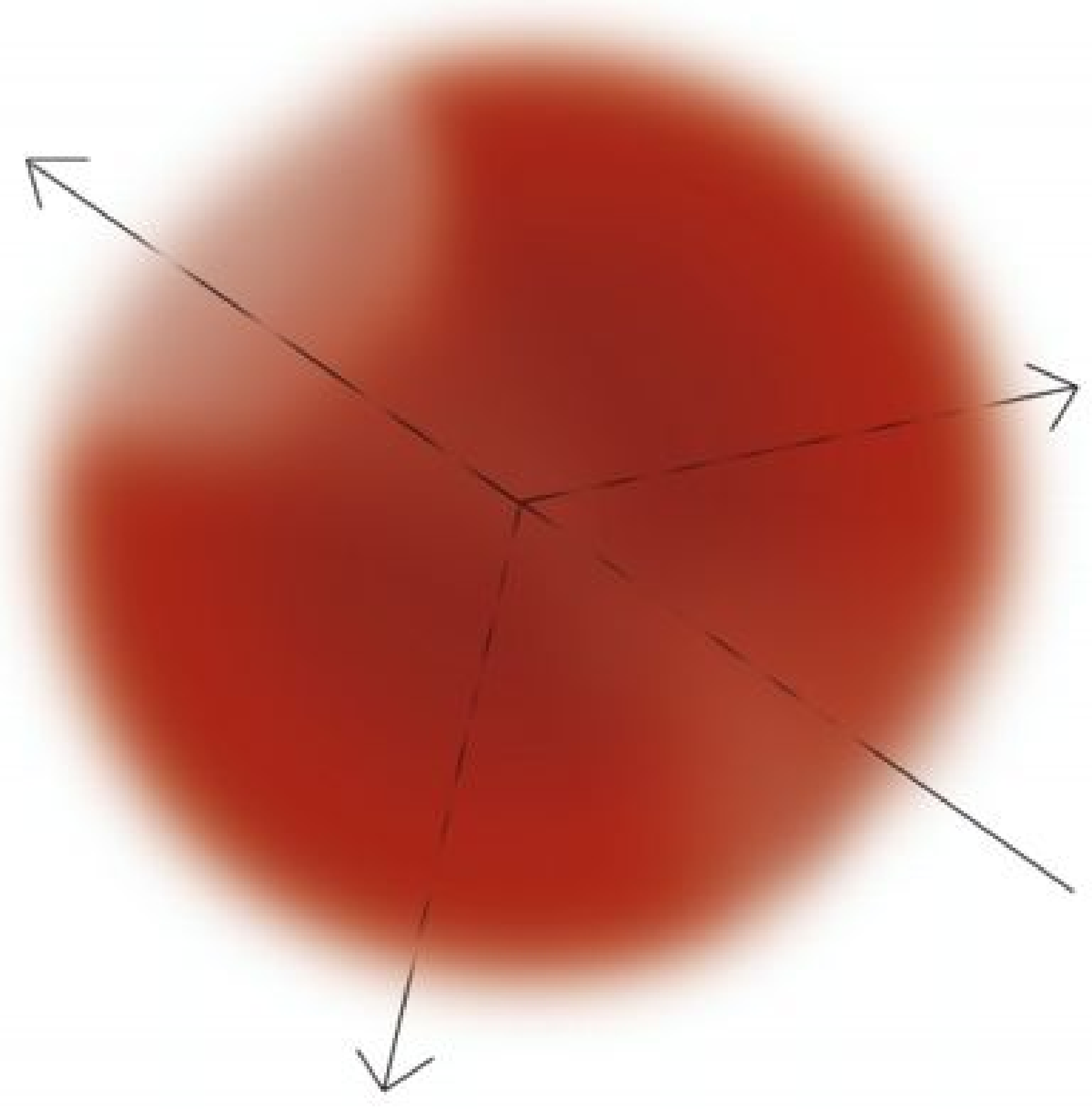}
			\includegraphics[height=4.15cm]{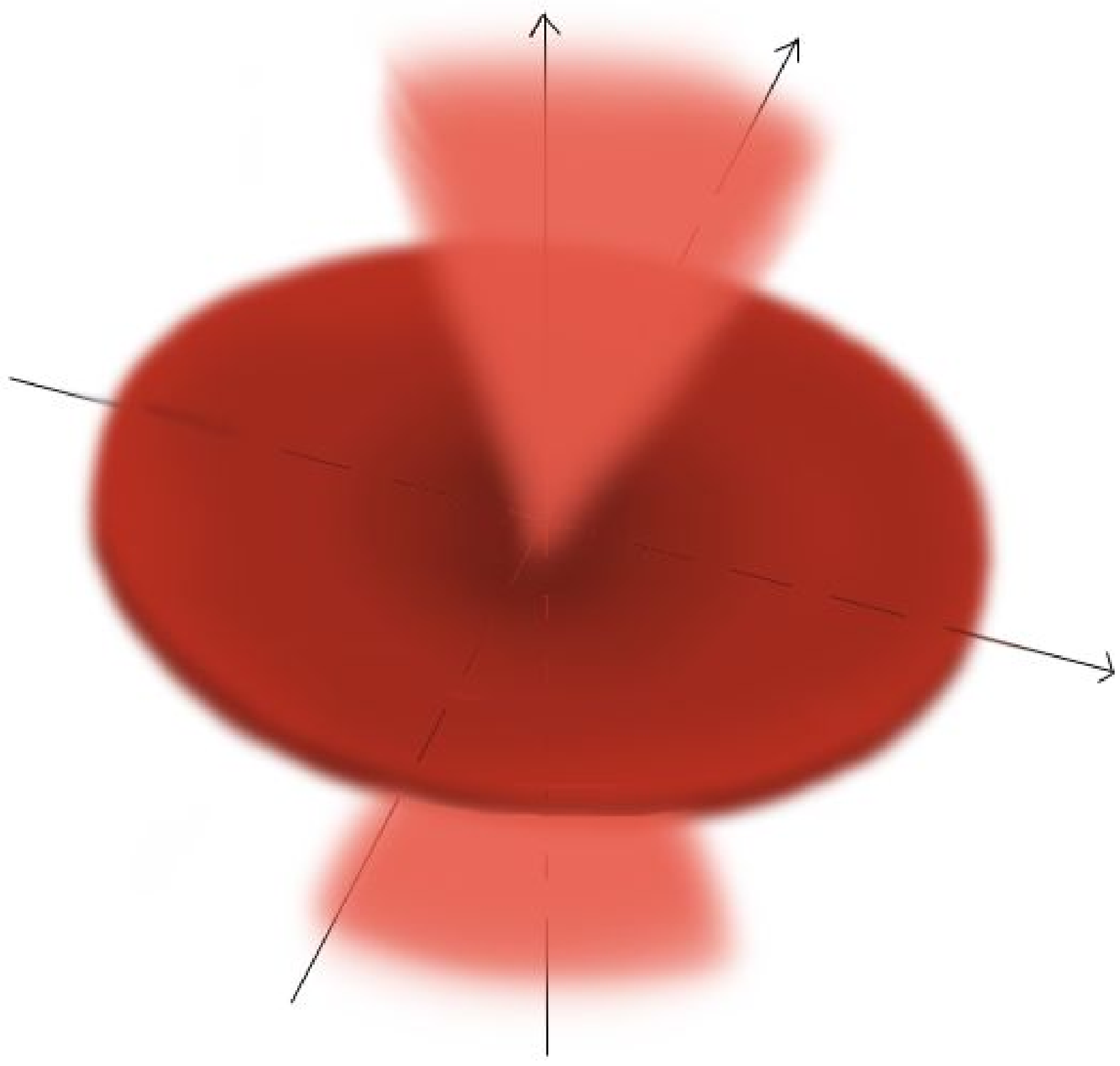}
			
		  \caption{"Toy" models used in order to fit the AMBER and MIDI measurements. The 4 upper pictures are a cut of the circumstellar disk in a plane defined by the observer line of sight and the stellar rotational axis (the observer is on the right for each picture), the corresponding projections into the sky-plane with over-plotted the interferometric data points from MIDI and AMBER are the central pictures whereas a "3D artist view" is plotted into the lower raw for each model.}
		   \label{schematic}
   \end{center}

\end{figure*}

\subsection{Envelope geometry}
The envelope extensions presented in Table \ref{Table_extension} are very sensitive to the sky-plane baseline orientation. This is particularly obvious from Fig. 2 where we have plotted $\alpha$ Arae's (unresolved star + uniform disc) model diameters as a function of the sky-plane baseline position.  In the following sections we present very simple models in order to constrain the geometry of the circumstellar envelope by fitting the data obtained (see Fig. 2). Note also that we present here models for the light distribution on the sky,  not "physical" models which will be presented hereafter.

\subsubsection{Equatorial disk perpendicular to the polarization}
Our starting point is the equatorial plane position of the circumstellar environment of $\alpha$ Arae deduced from the polarization measurement done by McLean et Clarke (1979) and Yudin et al. (1998). The polarization angle  measured is P.A.=172$\degr$ and any flattened envelope model should have a semi-major axis perpendicular to this direction, i.e. around 82$\degr$. 

\noindent In the following we consider simple axi-symmetric disk models, presented in Fig. \ref{schematic}, with 3 free parameters: 

\begin{enumerate}
\item The inclination angle ($i$) between the observer and the polar axis (i=0$\degr$ corresponds to pole-on). 
\item the opening angle ($\alpha$) of the disk.
\item the disk extension ($a$) in mas. 
\end{enumerate}

The shape of the projection of the disk onto the sky-plane depends only on the two first free parameters, i.e. $i$ and $\alpha$. For all these "toy" models the observer is on the right for the 4 upper pictures in Fig.~\ref{schematic}. The corresponding projections into the sky-plane with over-plotted the interferometric data points from MIDI and AMBER are the central pictures whereas "3D artist views" are plotted into the lower row for each model. 

The ratio between the projected major-axis ($a$) and the minor-axis ($b$) of the envelope is given by:

\begin{equation}
\frac {a} {b} = \frac {1} {\cos {i} + 2sin {\frac {\alpha} {4}} \sin{ (i - \frac {\alpha} {4})}}
\end{equation}

\begin{figure}	
\includegraphics[height=7.0cm]{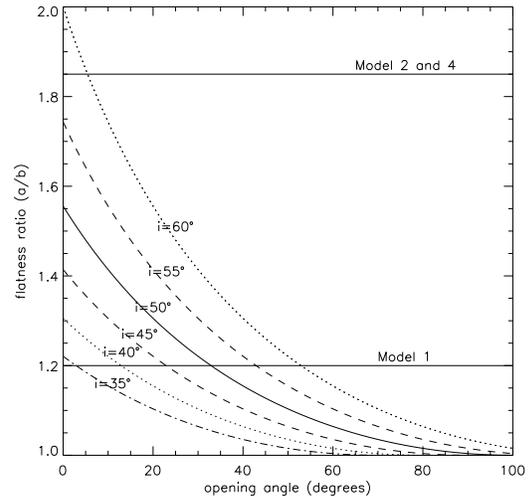}		
      \caption{Flatness ratio of the projected envelope versus opening angle calculted with our simple disc model for different inclination angle.  Horizontal lines show flatness from model 1 and model 2}
   \label{flatness}
\end{figure}

\noindent We assume that the major-axis of the envelope is oriented at P.A.= 82$^o$ (i.e. perpendicular to the polarization angle). Since B$_2$ is oriented at 81$\degr$ it is supposed to be a good estimate of the disk major-axis extension. In Fig.~\ref{flatness} we have plotted the ratio a/b as a function of the disk opening angle for different inclination angles between 35$\degr$ and 60$\degr$. This angle range was determined  following the inclination angle of 45$\degr$ for $\alpha$ Arae determined from the fit of the circumstellar  H$\alpha$, H$\beta$ and P$\beta$ emission lines in paper I. From Fig.~\ref{flatness} it is obvious that a precise determination of the inclination angle is mandatory in order to obtain an accurate opening angle estimation.\\

\noindent Our first disk model 1 with $i$=45 $\degr$, $\alpha$ = 25 $\degr$ and $a$=7.7 mas corresponds to a projected ellipse with a a/b ratio of 1.2. The agreement with the observed data is good excepted for the B$_3$ and B$_0$ baselines which present a smaller extension than predicted. In order to fit the data for the B$_3$ and B$_0$ baselines we define a model 2 with  $i$=57$\degr$, $\alpha$ = 2 $\degr$ and $a$=6.5 mas, i.e. smaller and very thin compared to model 1. In this latter case the a/b ratio is 1.85 which allows to fit the B$_2$ and B$_3$ baselines and MIDI data but not the B$_0$ and B$_1$ baselines. As shown in Fig.~\ref{flatness} model (2), with a very thin disk, is not compatible with an inclination angle of 45$\degr$ and we were obliged to use a larger inclination of 57$\degr$.  From  Fig.~\ref{flatness} we can see that for an inclination angle of 45$\degr$ the largest a/b ratio for a extremely thin disk ($\alpha \sim 0\degr$) is only 1.41, i.e. $\sqrt{2}$. Moreover, changing the inclination angle will change the shape of the H$\alpha$, H$\beta$ and P$\beta$
line profiles which were used to infer the value of 45$\degr$ obtained in paper I. Finally, using a simple disk model with various extensions, opening and inclination angles failed to fit simultaneously the AMBER and MIDI data especially for the B$_0$ and B$_3$  baselines which show a smaller extension along  the PA = 39$\degr$ and 52$\degr$ sky-plane orientation. 

\subsubsection{Polar-axis along the B$_3$ baseline orientation}
Since the extension obtained along the B$_0$ and B$_3$ baselines are the smallest ones we may consider that these baselines are close to the minor-axis of the envelope (i.e. along the stellar polar-axis).  Thus, our model 3 with $i$=45 $\degr$, $\alpha$ = 120$\degr$ and $a$=7.7 mas is a very thick disk as shown in Fig.~\ref{flatness} with the polar axis close to B$_0$ and B$_3$ but thick enough to go through the B$_1$ and B$_2$ data points. This model is in fact very similar to the one presented in paper I but in this case it maybe difficult to obtain a polarization large enough to be measured with a PA= 172$\degr$ since the major-axis of this "disk" is not perpendicular to the polarization direction.  Nevertheless, the polarization was measured in the visible, and since it originates from the inner part of the disk, this model cannot be totaly excluded.

\subsubsection{Equatorial disk + polar enhanced winds}
One of the shortcomings of the previous models is that they cannot reproduce simultaneously the two main envelope characteristics : the polarization angle at PA=172$\degr$ and the smallest  extension along the PA= 39$\degr$ whereas for the other AMBER baselines the disk is clearly more extended.\\

\noindent Our last simple model 4 try to take into account these observational characteristics by considering an equatorial disk oriented perpendicular to the polarization angle and flatened enougth to reproduce the difference between the extension measured along the B$_0$ and B$_2$ directions. Moreover, in order to fit the extensions measured for the B$_1$ baseline, we have added polar enhanced winds perpendicular to the disk as shown in Fig.~\ref{schematic}. Thus, seen into the sky plane it is also possible to fit all the interferometric MIDI and AMBER data points. Nevertheless, due to the fact that the disk is geometrically very thin ($\alpha$ = 2$\degr$) and for the same reasons already explain for our model 2 the inclination angle must be around 57 $\degr$. Moreover, the polar enhanced winds must be very extended in latitude (with an opening angle about 50$\degr$) and dense enough in order to fit the B$_1$ measurement.

\subsection{Conclusion about the ``toy story"}
Finally it seems that using very simple "toy" models we failed to reproduce the different extensions of the envelope as a function of the baseline projection onto the sky-plane. Thus, in the following section we will use as a starting point our model 4, i.e. geometrically thin disk + polar enhanced winds which was the one with the better agreement with both AMBER data and polarization direction. Moreover, the formation of geometricaly very thin disk, with a few degree for the opening angle  seems to be the best scenario up to now, at least for the central Keplerian disk around Be stars. 

\section{The SIMECA code: a brief description}
\label{secsimeca} 
In order to constrain the physical parameters of
the circumstellar environment of $\alpha$ Arae, we have used the
SIMECA code. This code, described in previous papers (see Stee \&
Ara\`ujo \citealp{Stee0}; Stee et al. \citealp{Stee1}; Stee \&
Bittar \citealp{Stee3}), has been developped to model the
environment of active hot stars.  SIMECA computes line profiles,
Spectral Energy Distributions (SEDs) and intensity maps, which can
directly be compared to high angular resolution observations.  The
envelope is supposed to be axi-symmetric with respect to the
rotational axis.  No meridional circulation is allowed. We also
assume that the physics of the polar regions is well represented
by a CAK type stellar wind model (Castor et al. \citealp{Castor})
and the solutions for all stellar latitudes are obtained by
introducing a parameterized model which is constrained by the
spectroscopic and interferometric data. The inner equatorial
region is dominated by rotation, therefore being quasi Keplerian.
The ionization-excitation equations are solved for an envelope
modeled in a 410{*}90{*}71 cube. 

The populations of the atomic levels are strongly altered by non-LTE
conditions from their LTE-values. For computation, we start with the
LTE populations for each level, and then compute the escape
probability of each transition, obtaining up-dated populations. By
using these populations as input values for the next step, we iterate
until convergence. The basic equations of the SIMECA code are given in
detail by Stee et al. \cite{Stee1}.

To take into account the photospheric absorption line, we assume
the underlying star to be a normal B3 V star with $T_{\rm eff} =
18\,000$\,K and $R=4.8 {\rm R}_{\odot}$ and synthesize the
photospheric line profiles using the SYNSPEC code by Hubeny
(Hubeny \citealp{Hubeny1}; Hubeny \& Lanz \citealp{Hubeny2}). The
resulting line profile is broadened by solid-body rotation and
might be further altered by absorption in the part of the envelope
in the line of sight towards the stellar disk.

\section{Using SIMECA for the modeling of $\alpha$ Arae circumstellar environement}
\label{bestmodel}
\subsection{Fit of the SED}
Thanks to the previous study we have now a good starting point for the estimation of the global geometry of the disk around $\alpha$ Arae.
Nevertheless, we want to use the SIMECA model already described in paper I and in section 3 in order to obtain a more physical scenario for this star.
Thus, in order to obtain a SED which can be used to constrain our model we have collected photometric measurements from the largest spectral range available, 
i.e. from UV to mid-IR.  We used UV measurements from Jamar \cite{jamar}, Thompson \cite{thomson},
U magnitude from Johnson \cite{johnson}, BVRIJHK magnitudes from Ducati \cite{ducati}, ISO data, MIDI SED measurements (2003) between 7-13 $\mu$m and IRAS data. Of course, using non contemporary data for a variable star may be questionable but this is a good starting point especially for the shorter wavelengths related
to the central star which is supposed to be stable, contrary to the circumstellar disk which may appear and vanish with a typical time-scale of a
decade and should modify drastically the IR excess of the SED. 

\begin{figure}
	\begin{center}
			\includegraphics[height=7.0cm]{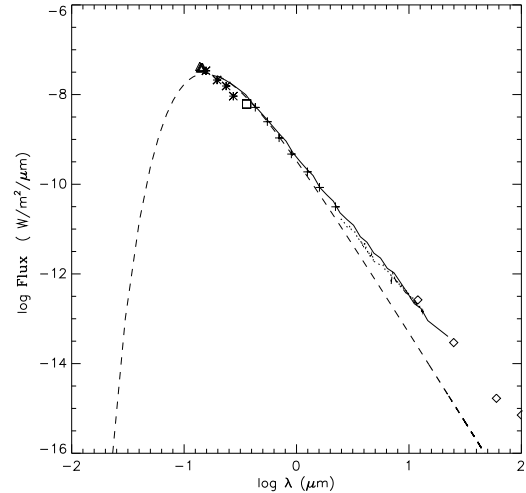}		       
      \caption{$\alpha$ Arae's Spectral Energy Distribution (SED) from various sources in the literature (see text). Dashed line: emission from the central star only assuming a black body with R$_\star$=4.8R$_\odot$, T$_{eff}$=18000K and d=105 pc. Fit of the SED with SIMECA taking into account the envelope free-free and free-bound contribution (plain line) between 0.3 and 20 $\mu$m}
   \label{sed}
   \end{center}
\end{figure}

The  SED we obtain from these various sources is plotted in Fig ~\ref{sed}. The SED is dominated by the emission of the central star for wavelengths smaller than 1$\mu$m which is assumed to be a black body with the following parameters from paper I: R$_\star$=4.8R$_\odot$, T$_{eff}$=18000K and d=105 pc.
We already discuss in paper I the fact that we were obliged to use a distance of 105 pc instead of the 74 pc obtained from Hipparcos measurements to fit the SED. Without considering any reddening, and keeping the Hipparcos distance, the radius of $\alpha$ Arae would be unrealistically low (below 3.5 R$\sun$) or the photosphere unrealistically cold (T$_{eff}$ $\sim$ 15000 K) thus we use the 105 pc determined from paper I in the following.
The  free-free and free-bound emissions from the envelope produce  an IR-excess and dominate the SED for wavelengths larger than $\sim$ 1 $\mu$m but the envelope remains optically thin.   This 
emission depends mostly on the number of free electrons and on the temperature law in the circumstellar envelope. As the envelope is almost fully ionized, we can consider that the global IR excess is only proportional to 
the mass of the disk. In Fig ~\ref{sed} we present our best fit of the SED using a temperature law in the envelope with T(r) $\propto$ r$^{-3/4}$ and a mass of the envelope of 4.1 ($\pm$ 0.5) 10$^{-10}$ M$_\odot$.

The mass of the disk in SIMECA depends on six parameters: three of them are related to the stellar mass loss: mass flux at the pole, C$_1$, and m$_1$. The three other ones are related to the envelope 
kinematics: terminal velocity at the pole, at the equator and m$_2$.

\noindent  The mass flux in SIMECA is given by:

\begin{equation}
\Phi(\theta)=\Phi_{pole}[1+(C1-1)\sin^{m1} (\theta)].
\end{equation}

\noindent where $m1$ is the first free parameter which describes the variation
of the mass flux from the pole to the equator, and C1 is the ratio between the equatorial and polar mass flux:

\begin{equation}
C1= \frac{\Phi_{eq}}{\Phi_{pole}}.
\end{equation}

\noindent The expansion velocity field is given by:

\begin{equation}
v_r(r,\theta)=V_o(\theta)+[V_\infty(\theta)-V_o(\theta)](1-\frac{R}{r})^{\gamma},
\end{equation}

\noindent We used  $\gamma$ = 0.86 which
is a typical value for early Be stars (Poe \& Friend \cite{poe}; Ara\'ujo \&
Freitas Pacheco \cite{araujo}; Owocki et al. \cite{owocki1}). 

\noindent with

\begin{equation}
V_o(\theta)=\frac{\Phi(\theta)}{\rho_{0}}=\frac{\Phi_{pole}[1+(C1-1)\sin^{m1} (\theta)]}{\rho_{0}}.
\end{equation}

\noindent The second free parameter $m2$ is introduced in the expression of the
terminal velocity as a function of the stellar latitude:

\begin{equation}
V_\infty(\theta)=V_\infty(pole)+[V_\infty(eq)-V_\infty(pole)]\sin^{m2} (\theta).
\end{equation}

\noindent {\bf The shape of the terminal velocity law as a function of the stellar latitude is plotted  Fig.~\ref{velocity}
as well as rotational velocity for various stellar radii.}\\

\begin{figure}
	\begin{center}
			\includegraphics[height=7.5cm, width=8.5cm]{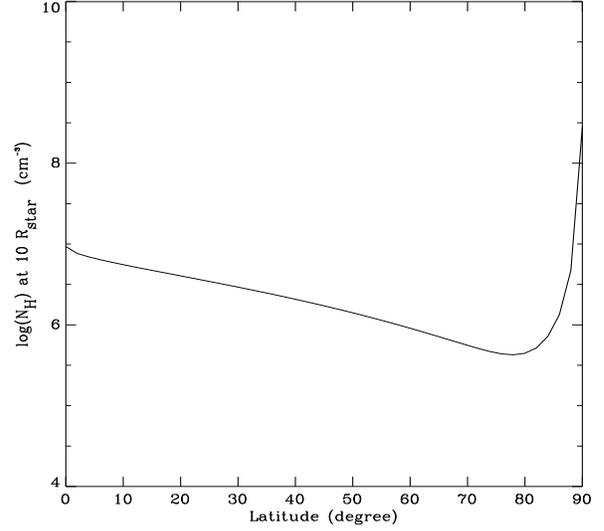}       
      \caption{Example of the SIMECA density distribution as a function of the stellar latitude from the pole ($\theta=0\degr$) to the equator ($\theta=90\degr$) at 10 R$_{*}$.}
   \label{density}
   \end{center}
\end{figure}

\begin{figure}
	\begin{center}
			\includegraphics[width=4.3cm]{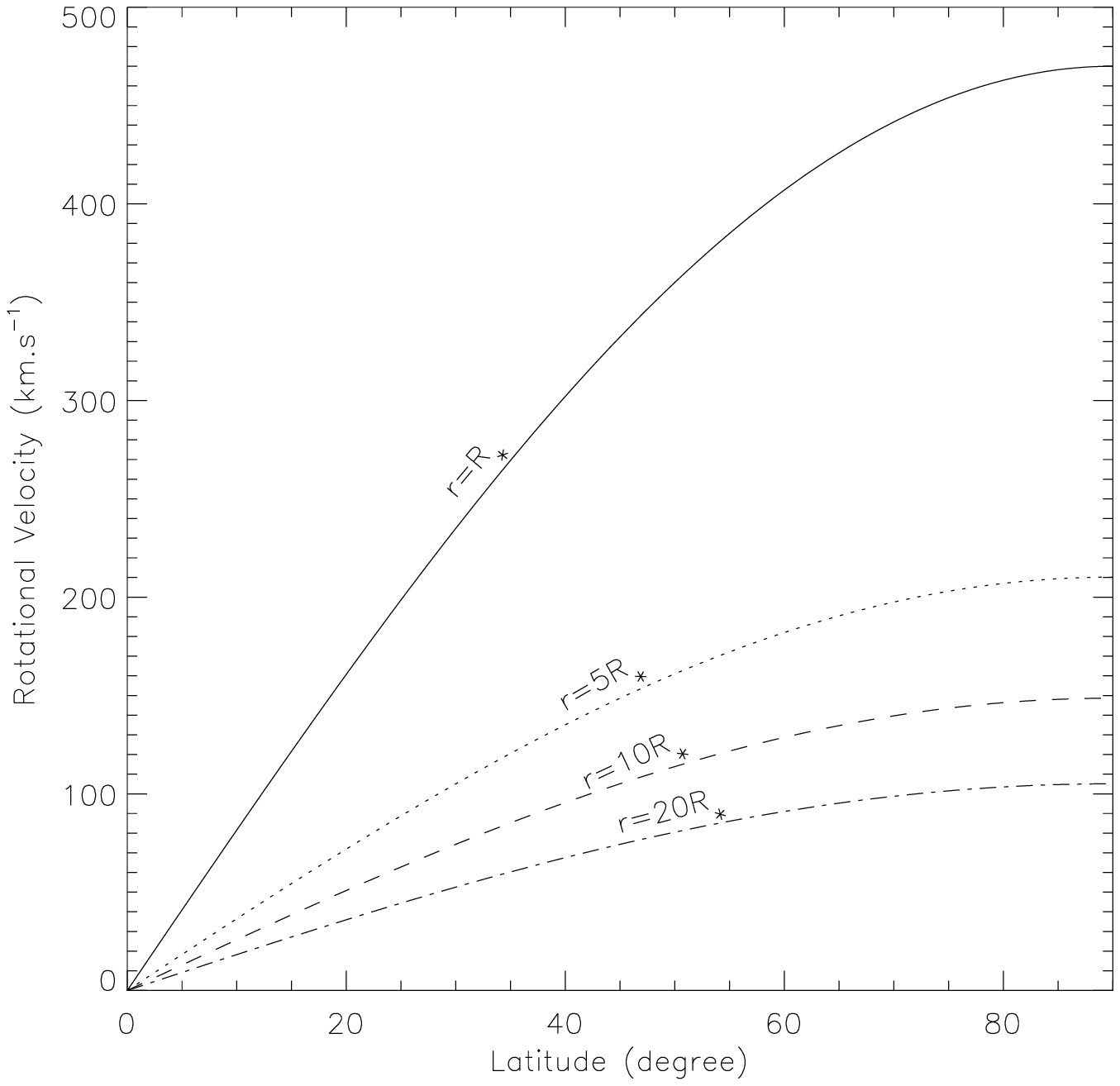}
			\includegraphics[width=4.3cm]{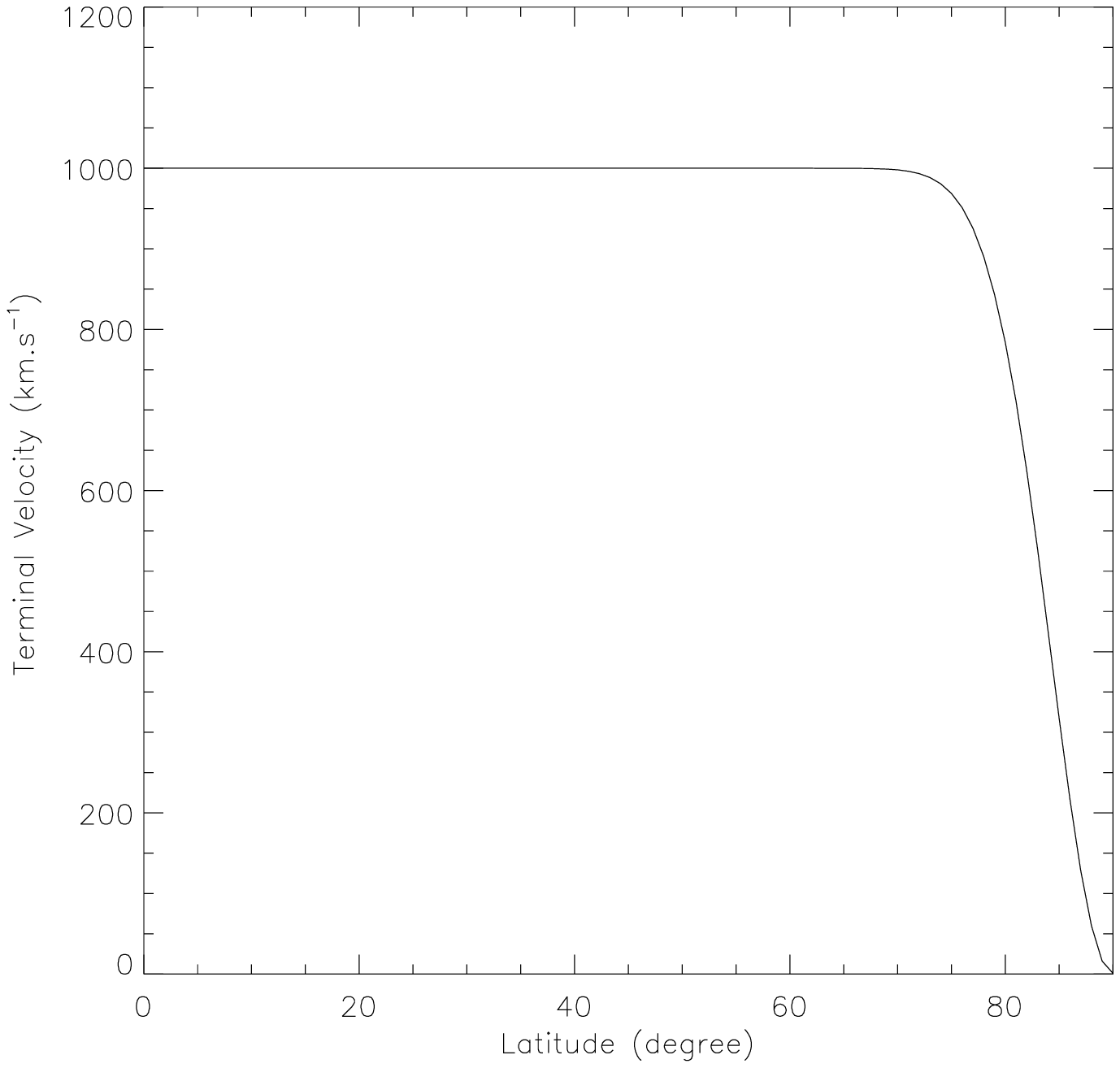}	         
      \caption{Left: rotational velocities as a function of the stellar latitudes from the pole ($\theta=0\degr$) to the equator ($\theta=90\degr$) at various stellar radii (over-plotted on the graph). Right:
      terminal velocity as a function of the stellar latitudes.}
      \label{velocity}
   \end{center}
\end{figure}

\noindent Finally the density distribution in the envelope is given by the equation of
mass conservation (see Fig.~\ref{density}):

\begin{equation}
\rho(r,\theta)=\frac{\Phi(\theta)}{{(\frac{r}{R})}^2 v_r(r,\theta)}.
\end{equation}

\noindent The mass of the disk is obtained by integrating Eq. (11). 
It is possible to obtain a combination of these six parameters that fit the SED but the solution may of course not be unique. In order to discriminate between all the possible
solutions  we need to put more constraints on our modeling, namely try to fit the visibility measurements in the lines and in the continuum for the various VLTI baseline orientations.

\subsection{Visibility modulus in the continuum}
\begin{figure}
	\begin{center}
			\includegraphics[width=8.5cm]{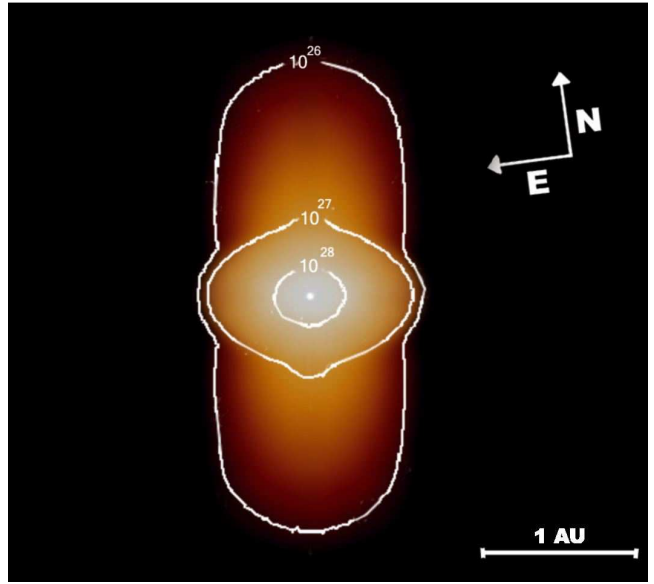}       
      \caption{Intensity map in the continuum at 2.15 $\mu$m obtained with SIMECA for our best model parameters. The inclination angle is 55$\degr$, the central bright region is the flux contribution from the thin equatorial disk whereas the smoother regions originate from the stellar wind. The brightness contrast between the disk and the wind is globally $\sim$ 30 but can reach 100 if you compare the inner region of the disk with the outer parts of the wind.}
   \label{map_continu}
   \end{center}
\end{figure}

In order to fit the visibility measurements described in section 4.2 we simulate a thin disk + polar enhanced winds with SIMECA with a dense equatorial  matter confined in the central region whereas
a polar wind is contributing along the rotational axis of the central star. Between these two regions the density must be low enough to reproduce the large visibility modulus (small extension) obtained for the B$_0$ and B$_3$ VLTI baselines.  Since $\rho$ $\propto$ $\frac{\phi}{v_{r}}$ (see Eq. 11) we can build a model satisfying the above conditions by tuning both the mass flux and the expansion velocity in the envelope. The expansion velocity at the equator should be very small, i.e. a few km.s$^{-1}$, whereas at the pole it can reach larger values up to 500-2000 km.s$^{-1}$. The density ratio between the equator and the pole must be around 10-100 to fit the data. The parameters we obtain for our best model are given in table ~\ref{best_model} and the 2.15 $\mu$m intensity map obtained in the continuum is plotted Fig ~\ref{map_continu}.  The largest discrepancies  between these parameters and those from paper I are that the star is now rotating close to its critical velocity (v=470 kms$^{-1}$, the disk is very thin but denser (by a factor 10)) and the expansion velocity is only 1 kms$^{-1}$. The continuum map is seen with an inclination angle of 55$\degr$, the central bright region is the flux contribution from the thin equatorial disk whereas the smoother regions originate from the stellar wind. The brightness contrast between the disk and the wind is $\sim$ 30 but can reach 100 if you compare the inner region of the disk with the outer parts of the wind. This density contrast implies that we use a C1 parameter smaller than 1, i.e. 0.03, in order to keep a sufficient density contrast equator/pole and a negligible equatorial expansion velocity
(1 kms$^{-1}$).\\

\begin{table}
{\centering \begin{tabular}{cc} \hline
parameter/result    & value \\
\hline
$v \sin i$& 375 km s\( ^{-1} \)\\
Inclination angle i & 55$\degr$\\
Photospheric density ($\rho_{phot}$)&1.0 10\( ^{-11} \)g cm\( ^{-3} \)\\
Equatorial rotation velocity & 470 km s\( ^{-1} \)  \\
Equatorial terminal velocity & 1 km s\( ^{-1} \) \\
Polar terminal velocity & 1000 km s\( ^{-1} \) \\
Polar mass flux & 7 10\( ^{-9} \)M\( _{\sun } \) year\( ^{-1} \) sr\( ^{-1} \) \\
m1 & 0.5 \\
m2 & 100.0 \\
C1 & 0.03\\
Mass of the disk & 4.1 10\( ^{-10} \)M\( _{\sun } \) \\
Mass loss & 1.3 10\( ^{-8} \)M\( _{\sun } \) year\( ^{-1} \)\\
\hline
\end{tabular}\par}

\caption{Best model parameters for the $\alpha$ Arae central star and its circumstellar environment
obtained from this work. Values for the other parameters that have not changed are listed Table~\ref{midi_model} and are from paper I.}
\label{best_model}
\end{table}

The fit of the visibility in the continuum basically constraints two parameters : the outer radius of the equatorial disk which will modify the visibilities for all the projected baselines  and the inclination angle which have a strong influence on the  flattening of the projected equatorial disk into the sky-plane.  The outer radius of the equatorial disk for our best model was set to 33 R$_\star$  which is larger compared to the 25 R$_\star$ found in Paper I but still in agreement with the 32 R$_\star$  obtained for the companion orbit.  Moreover, the inclination angle we obtain with our best model is 55$\pm5\degr$ which is also larger but in agreement with the inclination angle of 45$\pm5\degr$ found in paper I.

\subsection{Line profiles} 
In order to put additional constraints on our modeling we also try to fit  emission line profiles with SIMECA, namely H$\alpha$, H$\beta$ and Pa$\beta$. These lines were also used in paper I but following a quite different scenario: a nearly spherical expanding and rotating envelope versus a thin disk + polar wind in this paper.  Moreover, we also used in this paper the Br$\gamma$ emission line which is accessible to AMBER.  
We remind that the H$\alpha$ and H$\beta$ line profiles were observed in April 1999 with the HEROS instrument at la Silla (Chile), the Pa$\beta$ profile was recorded in August 13, 2003 at the Observatorio do Pico dos Dias (Brasil) and only the  Br$\gamma$ line was contemporary to our
AMBER interferometric observations.
Knowing that $\alpha$ Arae is a variable star exhibiting line variations with typical time scales of a few months, we will only use the global shape of these line profiles in order to constrain the kinematics within the disk. We will not try to fit  simultaneously the intensity of all these lines which is, moreover, not possible as shown in paper I. In our new model these emission lines originate from the dense and thin equatorial disk  and thus a fit of the shape of these lines will put strong constrains on the disk kinematics which is supposed to only rotate around the central star. Again, this is a quite different scenario compared to paper I where we used an expanding and rotating nearly spherical envelope to fit the lines.

\begin{figure}
	\begin{center}
		  \includegraphics[height=4.1cm]{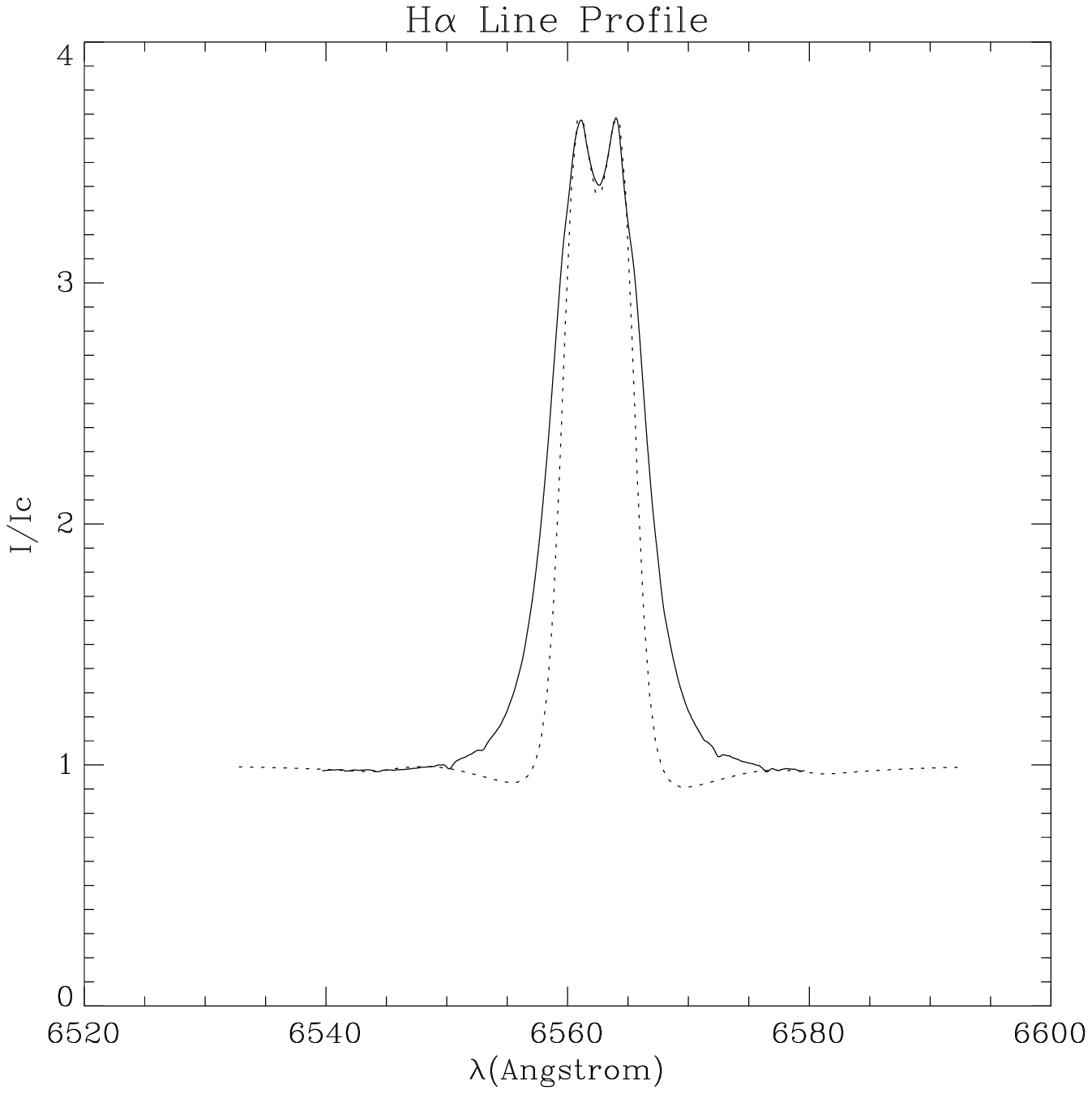}
		  \includegraphics[height=4.1cm]{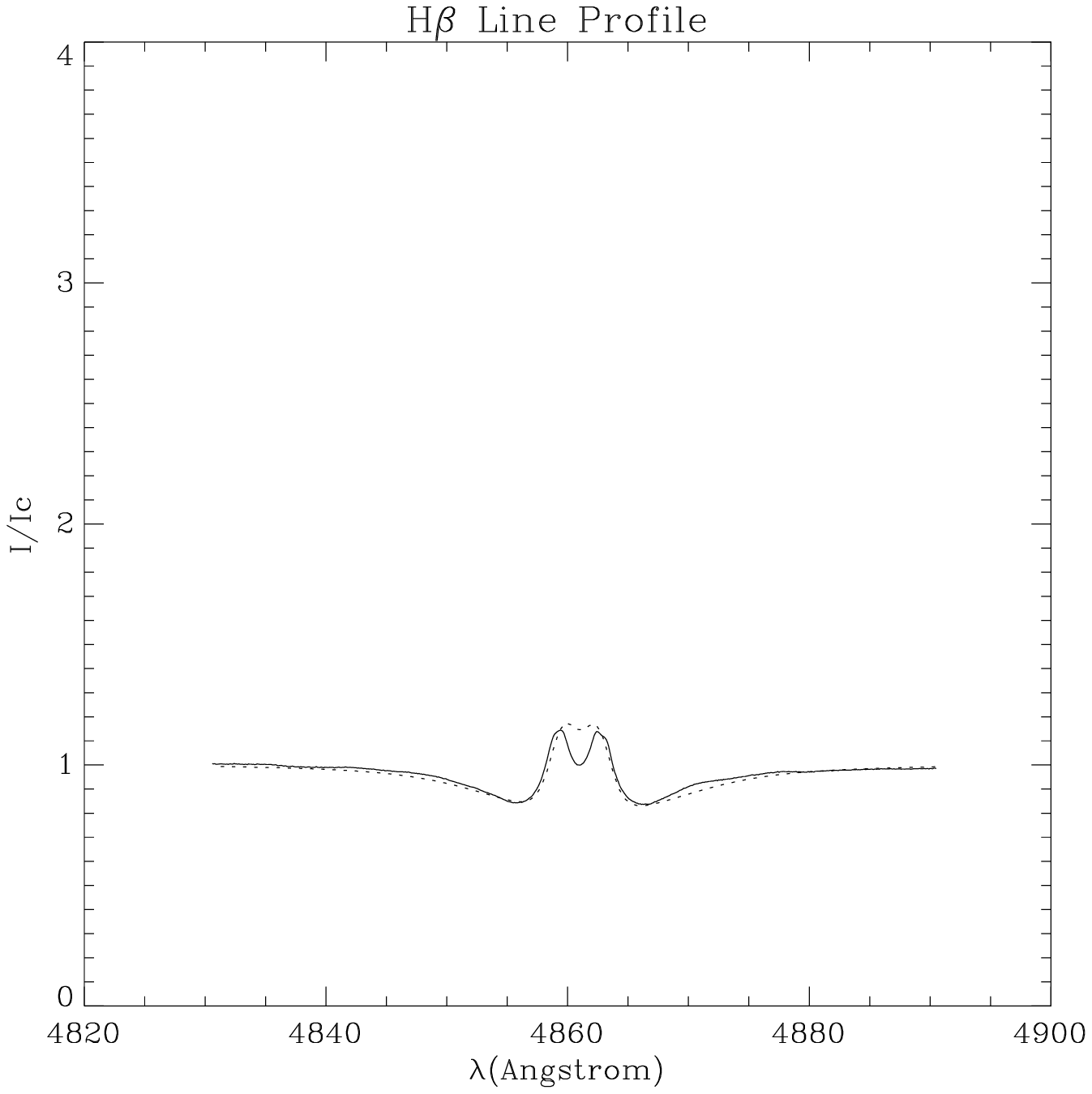}\\
		  \includegraphics[height=4.1cm]{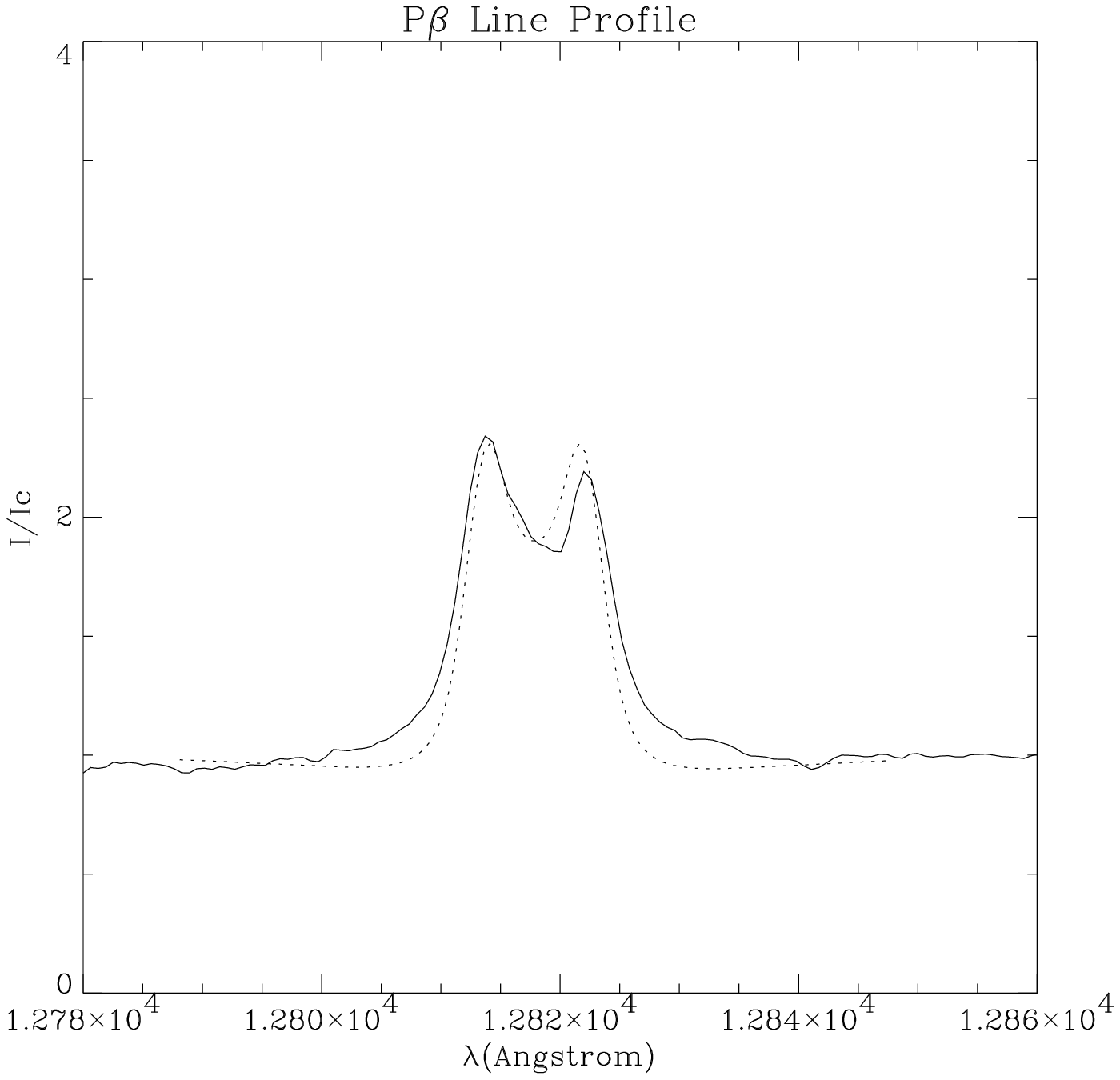}
		  \includegraphics[height=4.1cm]{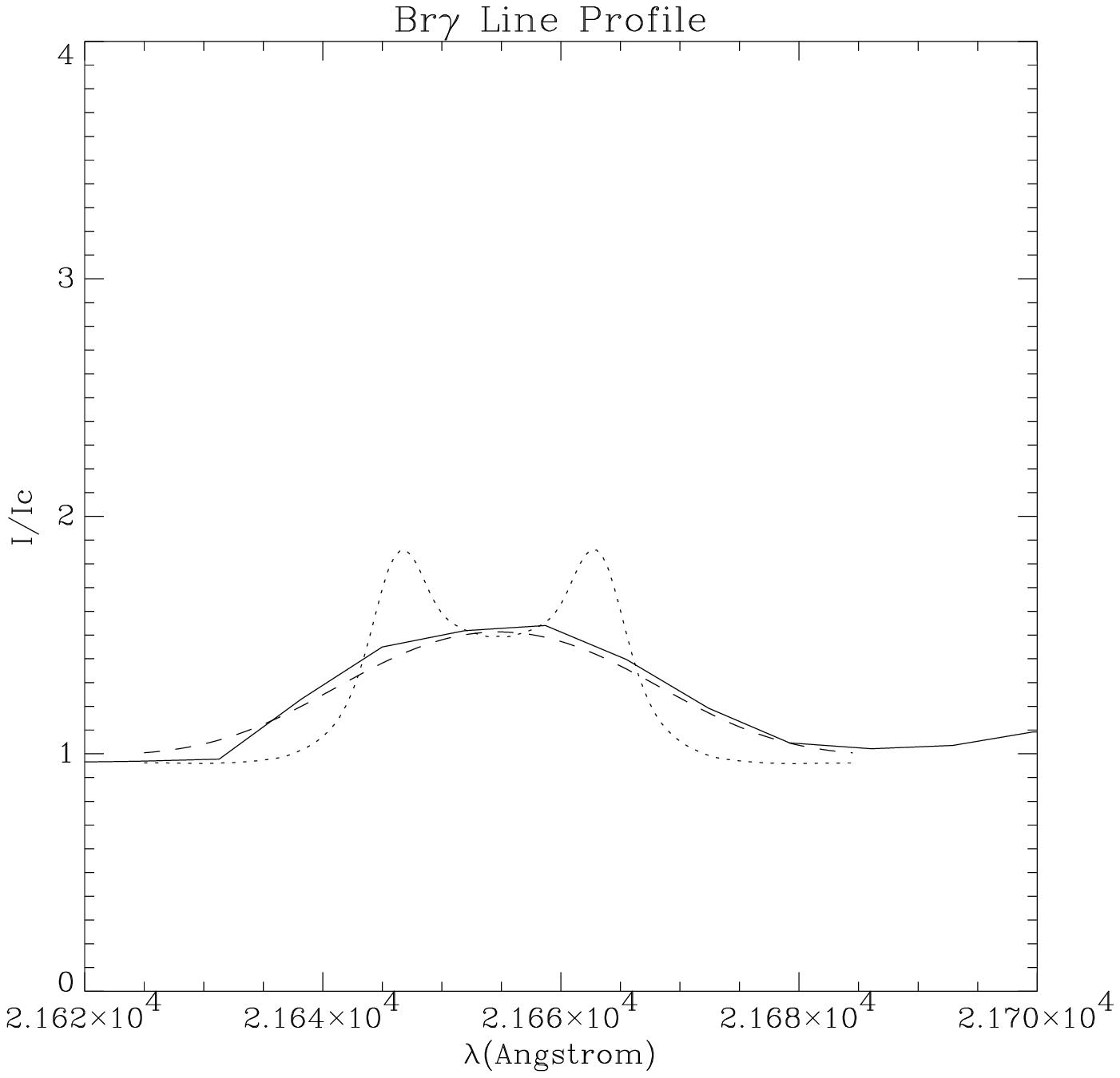}
      \caption{Line profiles modeled with the SIMECA code using a thin disk + polar wind scenario. Modeled profiles are dotted lines whereas observed ones are in plain line. For Br$\gamma$ we have convoluted the modeled profile (dotted line) obtain with SIMECA with a 15 $\AA$  gaussian, corresponding to the AMBER spectral resolution of 1500. The convoluted Br$\gamma$ profile we obtain is the dashed line superimposed with the observed one (plain line).}
   \label{profiles}
   \end{center}
\end{figure}

The observed Br$\gamma$ line profile is not showing a double-peaks structure due to the medium (1500) AMBER spectral resolution mode used for these first interferometric observation since the high (10000) AMBER spectral resolution mode was not available at this time. Nevertheless, we can see in Fig.\ref{profiles} that after a convolution with a 15 $\AA$  gaussian, corresponding to the AMBER spectral resolution, the agreement with the modeled profile (lower row, right profile, dashed line) and the AMBER ones (plain line) is satisfying.  For the other lines the fit of the peaks separation is in agreement with the keplerian rotation. As already discussed in paper I, we recall that between the 1999 observations (H$\alpha$ and H$\beta$ profiles), the 2003 (Pa$\beta$) and 2005 (Br$\gamma$) observations, we have decreased the density at the base of the stellar photosphere by 25 \% in order to fit the line intensity. 

Compared to paper I, the agreement between the modeled (dotted line) and observed (plain line) H$\alpha$ line profile in Fig.\ref{profiles}  is not as good. We are not able to reproduce the broad line wings which were mainly due to the nearly spherical expanding stellar wind used in our previous model. In our new scenario the geometrically thin and rotating equatorial disk produces a narrower H$\alpha$  line profile. In order to obtain larger line widths we should have taken into account multiple diffusion that occurs preferentially in the line wings as shown by Poeckert \& Marlborough \cite{poeckert2}. This is especially true for H$\alpha$ and Pa$\beta$ but less pronounced for H$\beta$ and Br$\gamma$. Nevertheless, we concentrate ourselves to the double-peaks separation which is very sensitive to the rotational velocity law used and as already mentioned, the agreement we obtain was better using a keplerian rotating law within the disk. 

\begin{figure}
	\begin{center}
			\includegraphics[width=9.0cm]{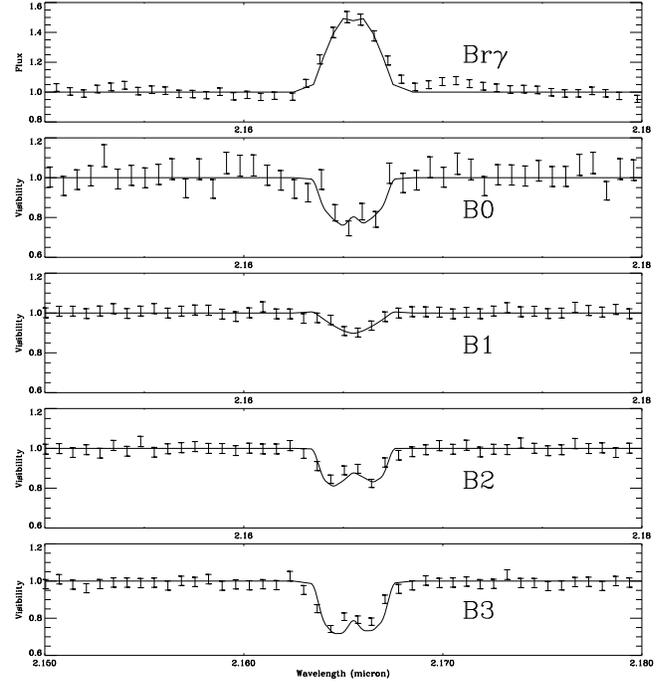}       
      \caption{Differential visibility across the Br$\gamma$ line profile for the B$_0$, B$_1$, B$_2$ and B$_3$ baselines. The first picture from the top is the Br$\gamma$ line profile. The plain line are the fits we obtain with SIMECA from our best model whereas the VLTI/AMBER data are the points with error bars.}
   \label{visimodulus}
   \end{center}
\end{figure}

\begin{figure}
	\begin{center}
			\includegraphics[width=9.0cm]{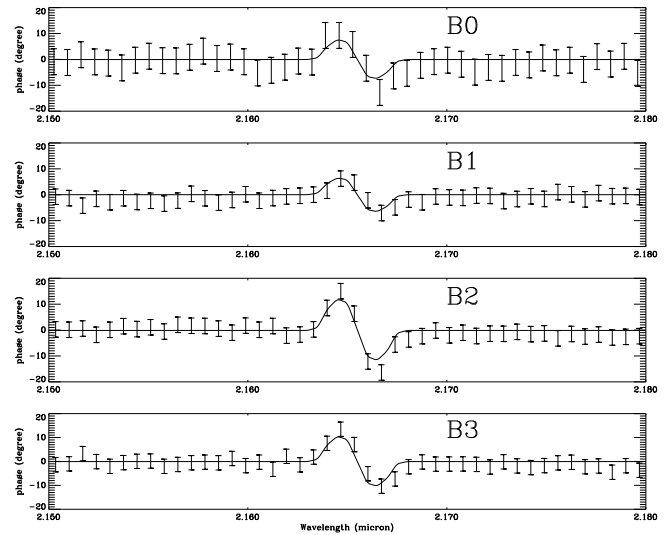}	         
      \caption{Phase of the visibility across the Br$\gamma$ line profile for the B$_0$, B$_1$, B$_2$ and B$_3$ baselines. The plain line are the fits we obtain with SIMECA from our best model whereas the VLTI/AMBER data are the points with error bars.}
   \label{visiphase}
   \end{center}
\end{figure}

\subsection{Differential visibility modulus across the Br$\gamma$ emission line}
\begin{figure*}[ht!]
	\begin{center}
			\includegraphics[width=18.0cm]{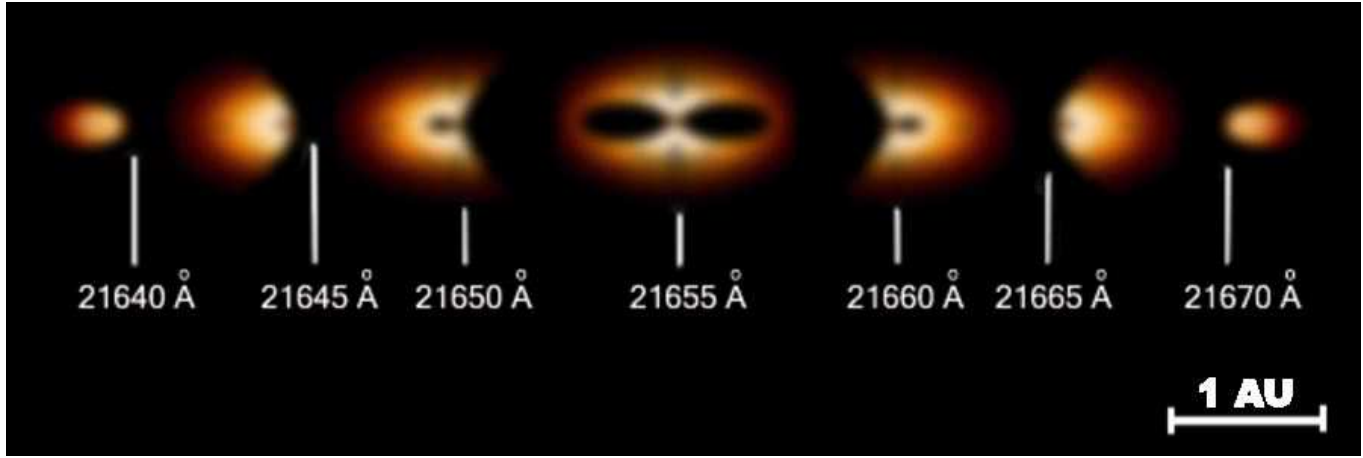}         
      \caption{Intensity maps across the Br$\gamma$ line profiles within spectral channels of 15 $\AA$ from which the differential visibility modulus and phase are estimated. Note that in order to increase the image contrast the central star and the continuum emission has been subtracted.}.
   \label{maps_raie}
   \end{center}
\end{figure*}

The differential visibilities curves are plotted Fig.~\ref{visimodulus} for the  B$_0$, B$_1$, B$_2$ and B$_3$ baselines. The agreement between the modeled visibility (plain-lines) and the VLTI/AMBER data is very good. Moreover, it was possible to reproduce the visibility decrease across the Br$\gamma$ line profile mainly due to the variation of the flux ratio between the unresolved star and the partially resolved circumstellar environment but also its shapes as a function of wavelength. For the B$_2$ baseline mainly along the equatorial disk direction, the  B$_0$ and B$_3$ baselines close to the equatorial disk, the curves, with a global "U" shape, present a visibility increase at the center of the line. This means that $\alpha$ Arae observed within a narrow spectral bandwidth of 15 $\AA$ appears smaller at the center of the emission line compared to its extension in the nearby emission line wings as seen in Fig.\ref{maps_raie}. On the contrary, for the B$_1$ baseline oriented along the polar-axis the visibility curve present a single "V" shape decrease at the line center. These effects are mainly due to the kinematics within the disk and are confirmed by the shape of the phase of the visibility as we will see in the following section. 

\begin{figure}
	\begin{center}
			\includegraphics[height=7.0cm]{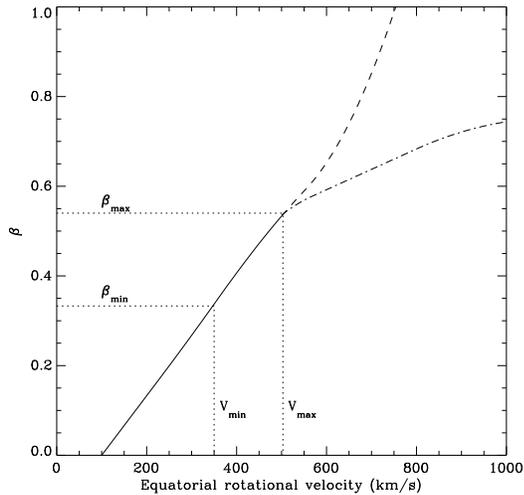}	         
      \caption{$\beta$ parameter as a function of the equatorial velocity taking into account the stellar photosphere deformation due to its fast rotation. The plain line corresponds to physically possible cases whereas the dashed line corresponds to rotational values larger than the "breakup" velocity. The dash-doted line also corresponds to unrealistic values but in this case we have not taken into account the stellar photosphere distortion larger than 1.5 R$_{\star}$ at the equator. v$_{min}$=351 kms$^{-1}$ corresponds to the minimum vsini found in the literature for this star, i.e 288 kms$^{-1}$, by Uesugi \& Fukuda \cite{uesugi} with i=55$\degr$.}.
   \label{vrot_beta}
   \end{center}
\end{figure}

\subsection{Differential phase across the Br$\gamma$ emission line}
The shape of the differential phase across the Br$\gamma$ line profile is related to the photocenter displacement of the object
as a function of wavelength and is very sensitive to the $\beta$ law used in the modeling as already shown by Stee \cite{Stee5}.
For our best model we used a keplerian rotation within the disk and the resulting visibility phases as a function of wavelength for the 
B$_0$, B$_1$, B$_2$ and B$_3$ baselines are plotted Fig.~\ref{visiphase}. Again, the agreement between the modeled phases and the 
VLTI/AMBER data is very good which is the {\bf first direct evidence of the keplerian rotation within a Be circumstellar disk}. Nevertheless, in order to know if this agreement is unique we have tested different models with different disk rotational velocities following:

\begin{equation}
 {v_{\phi}}(r,\theta)\propto{sin\theta} {\Bigl ( {R \over r} \Bigr )^{\beta} },
\end{equation}

with $\beta$ between the constant rotation ($\beta$ = 0) and the angular momentum conservation ($\beta$ = 1.0), respectively
$\beta$=0, 0.3, 0.4, 0.45, 0.5, 0.55, 0.7 and 1. Nevertheless, if you modify the $\beta$ law and still want to fit the double-peaks of the Br$\gamma$ emission line profile you need to simultaneously modify the equatorial rotational velocity of the star since the inner part of the disk,
supposed to be in contact with the star's photosphere, rotates at the same velocity. Thus we were obliged to take into account the shape of the star photosphere and its distortion due to its fast rotational velocity as shown by Domiciano de Souza et al.\cite{domiciano} for the Be star Achernar which was rotating close to its critical velocity and thus exhibit a flattened photosphere with an equatorial vs polar radius ratio of 1.5. 
In Fig.~\ref{vrot_beta} we have plotted the "beta" law versus the equatorial rotational velocity  that fit the double-peaks of the 
Br$\gamma$ line emission line profile, assuming a purely circumstellar rotating disk without expansion. This figure shows that if we assume that the stellar rotation is between v$_{min}$ corresponding to the smaller measured vsini=288 kms$^{-1}$ by Uesugi \& Fukuda \cite{uesugi}, with i=55$\degr$, i.e. v$_{min}$= 350 kms$^{-1}$ and v$_{max}$, corresponding to the critical velocity for this star, i.e. v$_{max}$=503 kms$^{-1}$; the $\beta$ parameter must lies between 0.33 and 0.54. This last value is in better agreement with the upper value of the measured vsini=375 kms$^{-1}$ by Bernacca \& Perinotto \cite{bernacca} with i=55$\degr$, i.e. v=457 kms$^{-1}$ at the equator which may be a good indication that {\bf $\alpha$ Arae is rotating very close to its critical velocity}.\\

\begin{figure}
	\begin{center}
			\includegraphics[width=7.0cm]{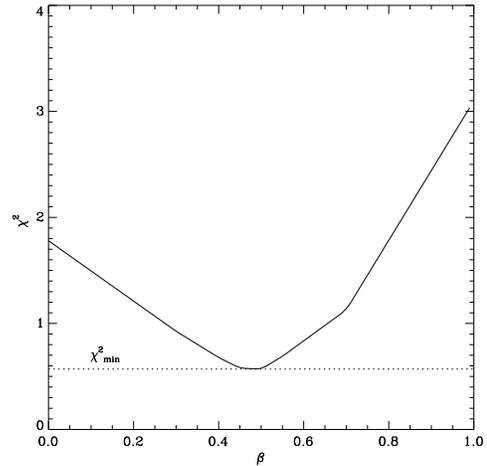}        
      \caption{$\chi ^{2}$ from our models and the AMBER data plotted as a function of the $\beta$ parameter. $\chi ^{2}_{min}$ is the minimum value obtained for a model with $\beta$=0.48 which is very close to the Keplerian rotation ($\beta$=0.5). }
   \label{chi2}
   \end{center}
\end{figure}

Moreover, in order to be more quantitative we have computed the total $\chi ^{2}$ from our best model as a function of the $\beta$ parameter. The results are plotted Fig.\ref{chi2} and we clearly see that the better agreement ($\chi ^{2}_{min}$) is obtained for a rotating law very close to the Keplerian rotation, i.e. $\chi ^{2}_{min}$=0.48 $\pm$0.04. 

\begin{figure*}[ht!]
	\begin{center}
	    \includegraphics[width=7cm]{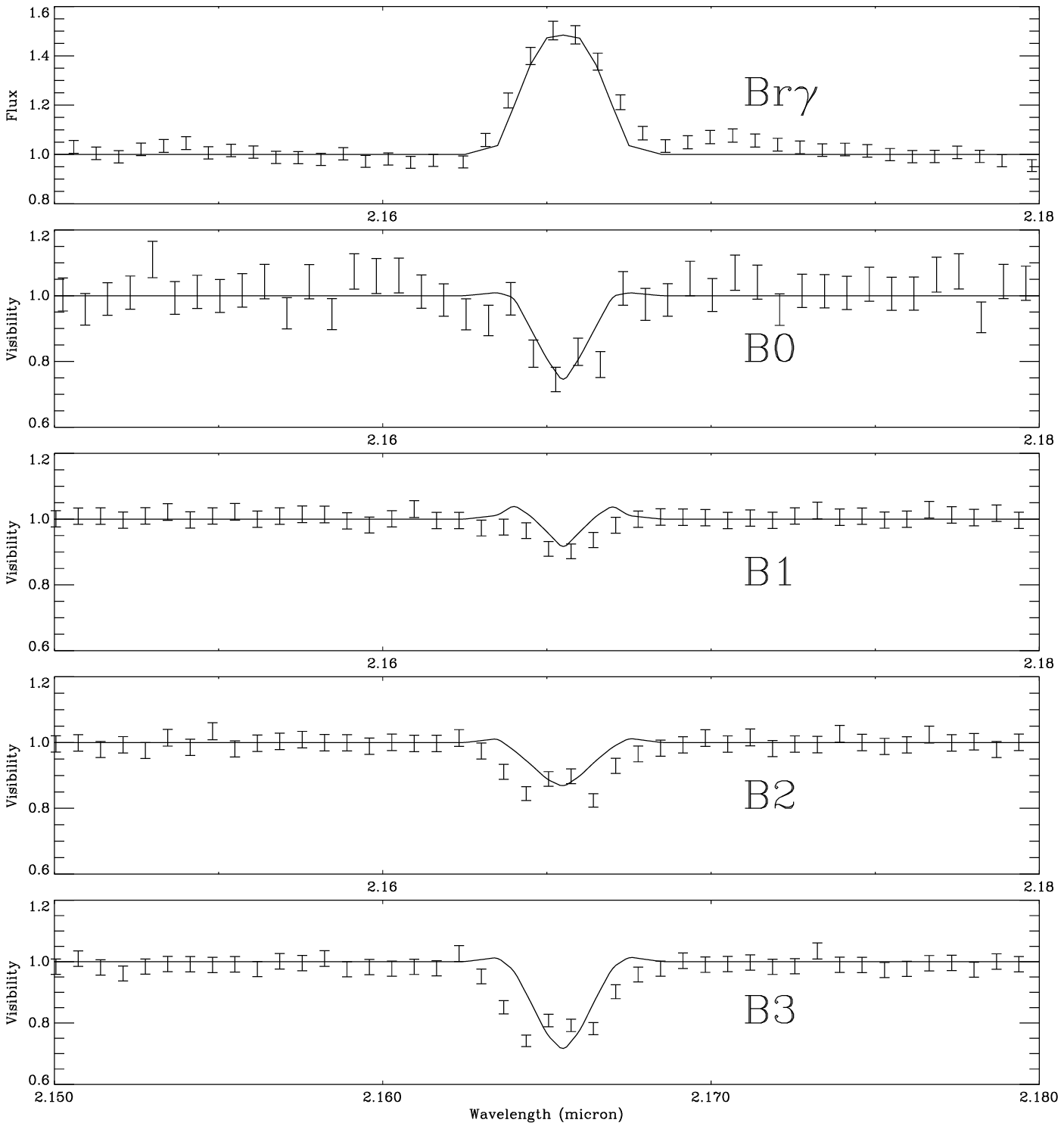}	
	    \includegraphics[width=7cm]{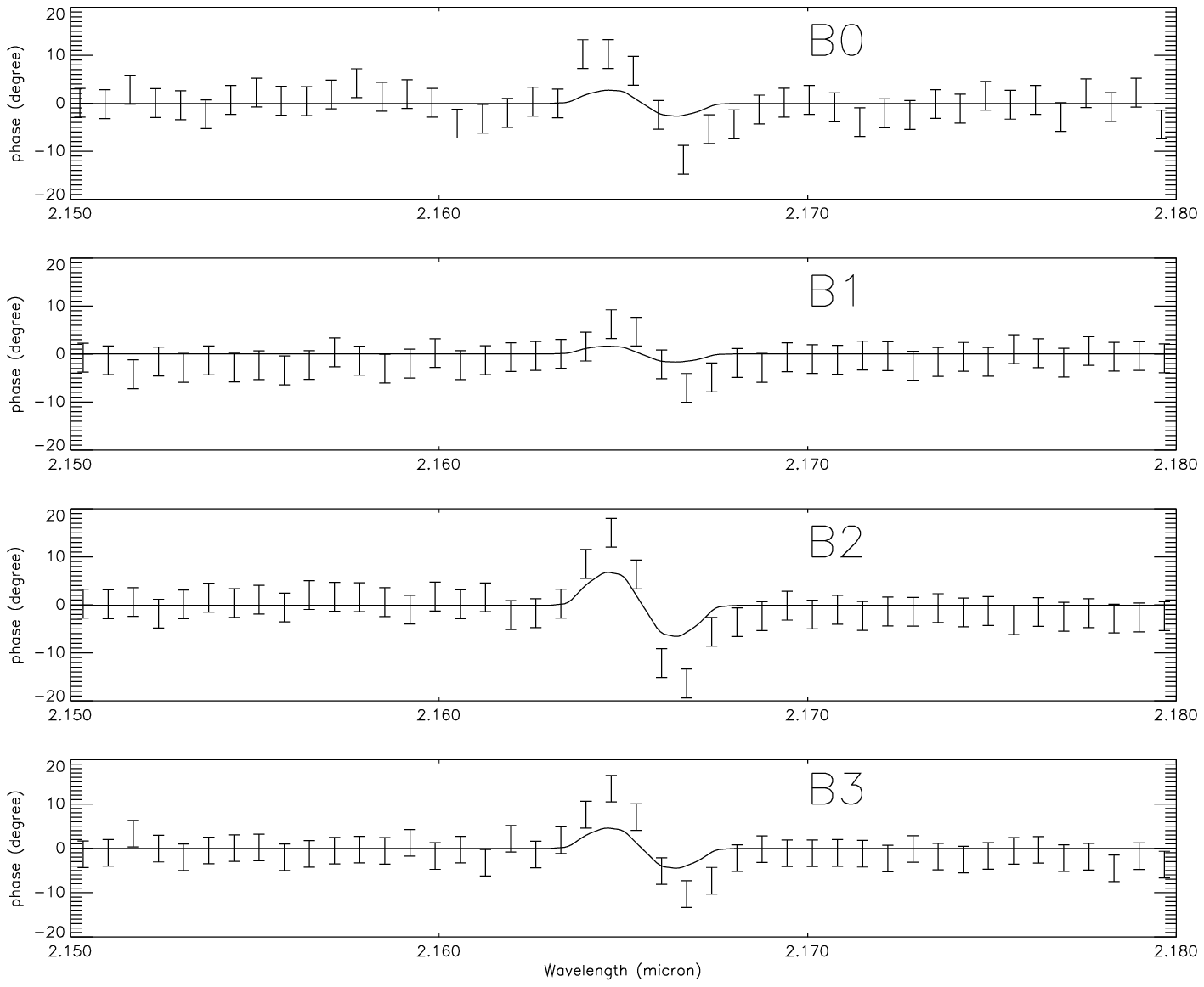}
		
      \caption{$\alpha$ Arae differential visibilities (left) and phases (right) across the Br$\gamma$ line profiles for the B$_0$, B$_1$, B$_2$ and B$_3$ baselines. The plain line are the fits we obtain with SIMECA from a rotating and expanding scenario decribed in section \ref{other_scenarios} whereas the VLTI/AMBER data are the points with error bars }
   \end{center}
  \label{model_exp+disc}
\end{figure*}

\begin{figure*}[ht!]
	\begin{center}	   
			\includegraphics[width=7cm]{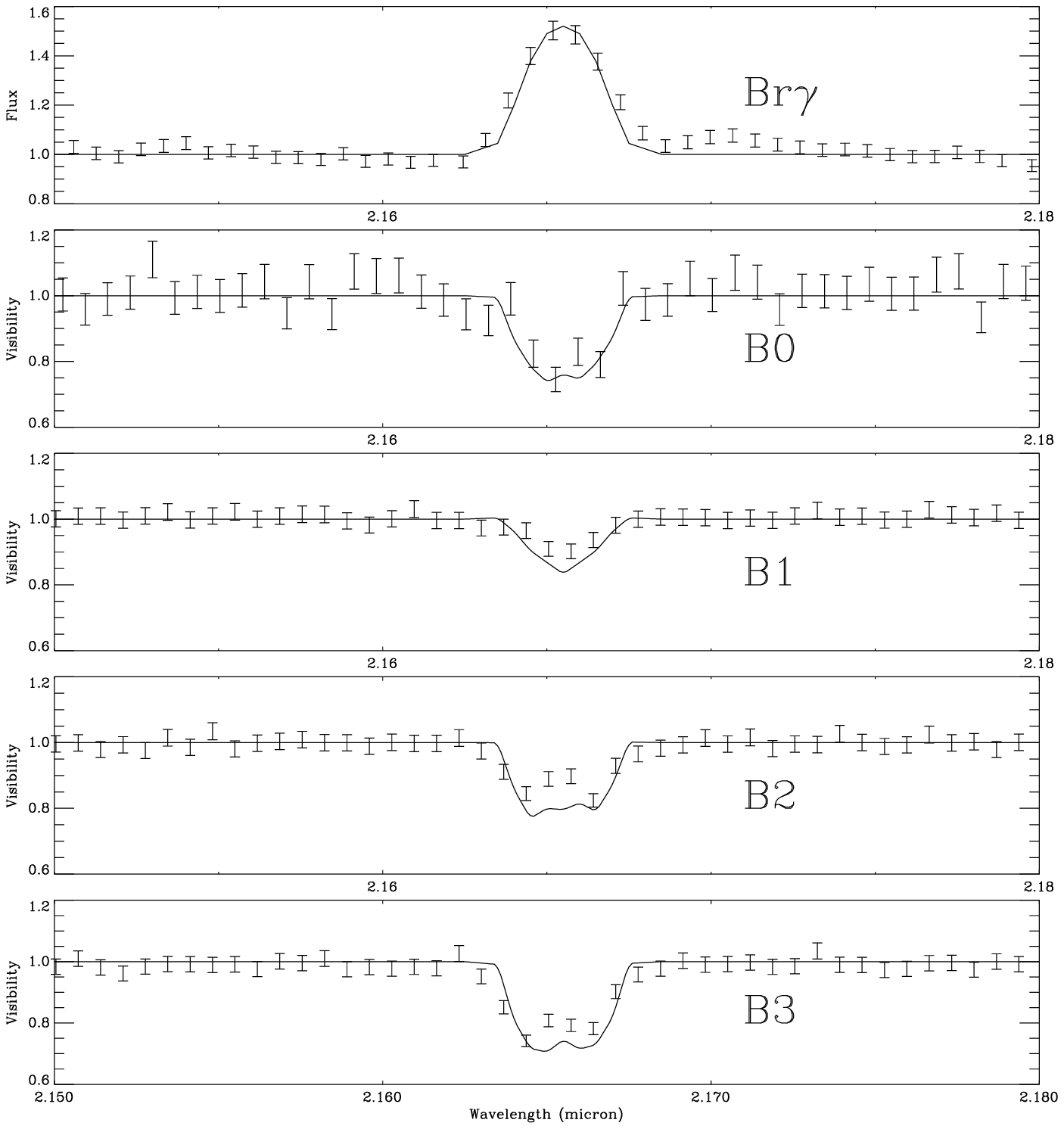}			\includegraphics[width=7cm]{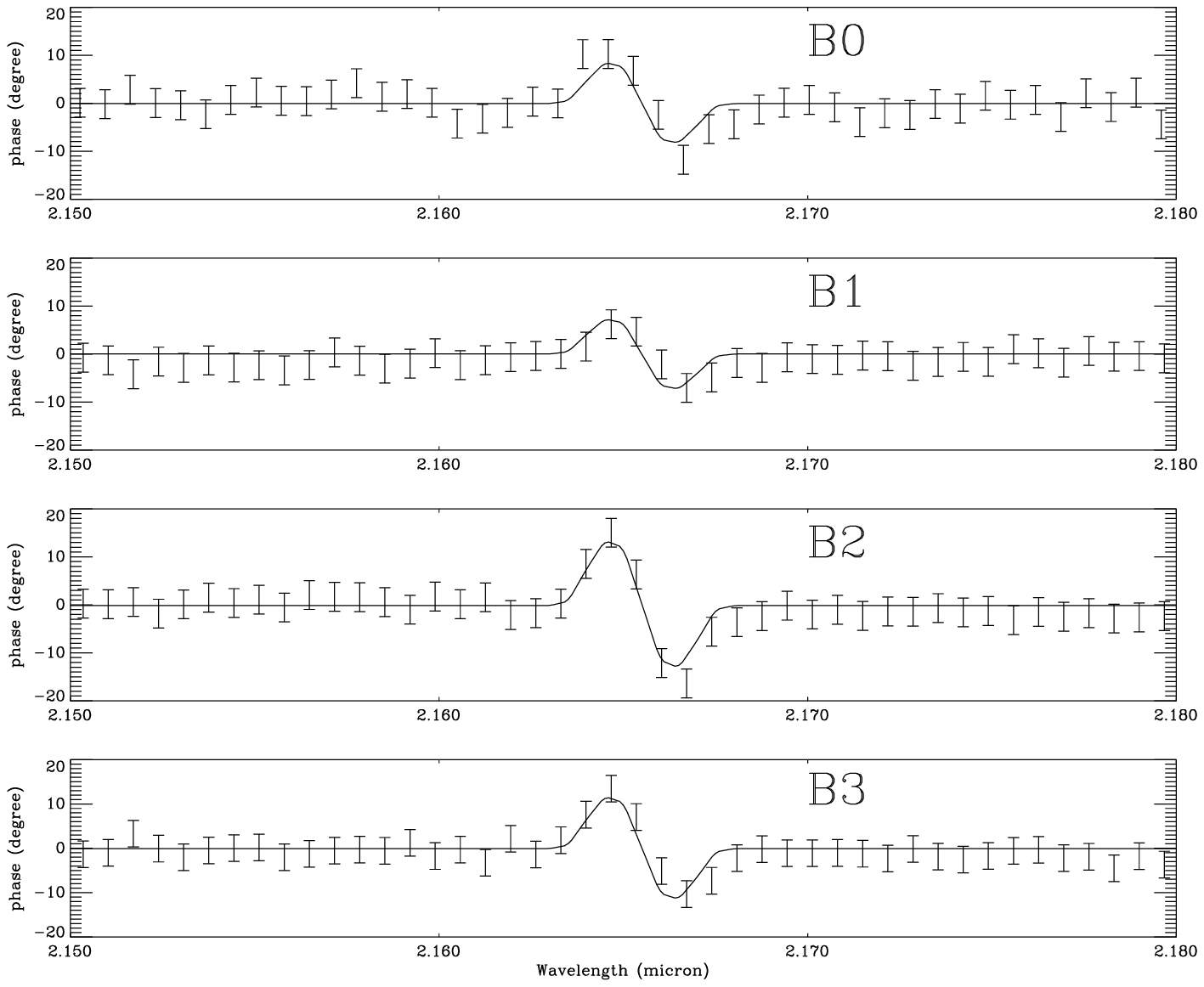}
      \caption{$\alpha$ Arae differential visibilities (left) and phases (right) across the Br$\gamma$ line profiles for the B$_0$, B$_1$, B$_2$ and B$_3$ baselines. The plain line are the fits we obtain with SIMECA from the model 3 decribed in section 4.3.2 and section  \ref{other_scenarios} whereas the VLTI/AMBER data are the points with error bars }
   \end{center}
  \label{model_kep+sphere}
\end{figure*}

\subsection{Is the geometrically thin and rotating disk scenario unique ?}
\label{other_scenarios}
Since our Keplerian disk + polar enhanced winds scenario seems to successfully reproduce all the available observables (i.e. photometric, spectroscopic, polarimetric and interferometric) we may wonder if this scenario is really unique, especially due to the number of parameters used in the SIMECA code. Thus, we have tested two other scenarios, the first one is based on the same global geometry of our best model, but we have now
added an expansion component produced by an equatorial stellar wind.\\

\noindent Since the peaks separation of the line profiles depends on both components of the velocity field, i.e. expansion and rotation, we have to decrease the rotational velocity used in our best model to keep the same observed peaks separation. Remembering that the lowest possible value of the rotational velocity is 351kms$^{-1}$ we can only use a maximum value of 100km$^{-1}$ for the expansion equatorial terminal velocity. The corresponding differential visibilities and phases we obtain are plotted Fig.~\ref{model_exp+disc}.  In this case, the decrease of the differential phase amplitude variations is due to the smaller value used for the rotating component. If the envelope was only in spherical expansion due to a spherical stellar wind, the phase variation
across the Br$\gamma$ line would be zero since there is no photocenter displacement for a spherically symmetric velocity field. On the contrary, if you have a purely rotating disk, the photocenter displacement will follow the projection of the iso-velocities regions and will produce a typical ``S-shape" as shown Fig.~\ref{visiphase}. Thus, a decrease of the rotational velocity field component regarding the expansion component produces globally a smaller differential phase amplitude variation shown Fig. 14. The amplitude of the differential visibilty across the line remains almost
the same since it is less sensitive to the kinematics within the disk but rather to a global geometric change of the circumstellar environment and to a change in the star versus envelope flux ratio  in the line and the continuum which remains unchanged by modifying the velocity fields. Nevertheless, the shape of the differential visibility is strongly modified and presents now a ``V" shape as already mentioned in section 6.4. This shape variation is due to the fact that the expansion versus rotation ratio is increasing in the equatorial region and the difference between the pole versus equator kinematics is less pronounced when adding an equatorial expansion component. This analysis exhibit the importance of spectrally resolved interferometric measurements for the study of the kinematics within circumstellar disks.\\

The second scenario is based on a quasi-spherical model as the model 3 already described in our ``toy story" section. The kinematic used is very close to our thin disk + polar enhanced winds model, (i.e. Keplerian rotation with a 1km$^{-1}$ equatorial terminal velocity) but in this later case the disk is rather geometrically thick. Again, the corresponding differential visibilities and phases we obtain are plotted Fig. 15.  Since the kinematics remains mostly unchanged the fits of the differential phases are as good as for our best model but since the geometry of the disk is different the agreement with the differential visibilities is not as good. Again, thanks to the same disk kinematics the shape of the differential visibilities is very similar but their amplitudes are not well reproduced. Nevertheless, these differences remain very small for an inclination angle of 55$\degr$ and it is very hard to put an upper limit for the disk opening angle using this method, especially  regarding the actual errors on the AMBER absolute calibrated visibilities. On the other side this study clearly shows that the equatorial region is a Keplerian rotating disk rather than an expanding wind.

\section{Discussion}
This study point out three important results touching lively debated issues:
\subsection{Keplerian rotation}
There were already some  indications that the disk may follow the Keplerian rotation by other theoretical studies, for instance the results obtained by Hanuschik~\cite{Hanuschik1} regarding
shell lines produced within a Keplerian disk in hydrostatic equilibrium. Hanuschik~\cite{Hanuschik2}; \cite{Hanuschik3} also investigate the geometrical structure of the emitting part of circumstellar envelopes around Be stars and found a good agreement with spectroscopic data using a thin disk in vertical hydrostatic and horizontal centrifuginal equilibrium, similar to a Keplerian accretion disk. These results
were confirmed by Rivinius~\cite{Rivinius1} presenting high resolution echelle spectra for 6 B-type stars supposed to be seen edge-on and in good agreement with Hanuschick's models for the formation of shell lines in circumstellar disks with Keplerian rotation. In a more recent paper Rivinius~\cite{Rivinius1} propose a scenario where the disk is no more a completely stationary structure but rather a succession of outbursts which may form rings. But even within this scenario, part of the ejected matters attains sifficiently high angular momentum to form a roughly Keplerian disk, at least for the immediate times of outbursts. Finally, another indirect argument in favor of a Keplerian disk is the success of the global oscillation modeling already outlined in the paper review by Porter \& Rivinius~\cite{porter}. Thus, our results
may be the way to put a final exclamation mark regarding the widely accepted fact that {\bf circumstellar disk around Be stars are in Keplerian rotation}.

\subsection{Stellar rotation}
As already mentioned in section 6.5 we found that $\alpha$ Arae must be {\bf rotating very close to its critical velocity} since we obtain $\frac{v_{rot}}{v_{crit}} \sim 91 \%$. This value is far above the conservative estimates of $\sim$ 75 \% usually found in the literature for Be stars. The fact that Be stars may be rotating much closer to their critical velocities than it is generally supposed was already outlined by Townsend et al.~\cite{townsend} and Owocki~\cite{owocki3}. This nearly critical rotation has quite profound implications for dynamical models of Be disk formation and may be the clue for the Be phenomenon. It may bring sufficient energy to levitate material in a strong gravitational field or at least help other physical processes such as pulsation or gas pressure  to provide sufficient energy and angular momentum to create a circumstellar disk. Moreover, observational evidences of this nearly critical rotation are growing such as the results obtained by Domiciano et al.~\cite{domiciano} using interferometric VLTI/VINCI data of Achernar.  They measured a rotationally distorted  photosphere with an apparent oblateness of 1.56 which cannot be explained using the classical Roche approximation. This scenario follows the original picture by Struve~\cite{struve} of a critically rotating star, ejecting material from its equatorial regions. 

\subsection{Polar wind enhancement}
Our interferometric measurements are evidencing {\bf a polar wind enhancement} (see Fig.~\ref{map_continu}) which was already predicted for almost critically rotating stars. For instance, Cranmer \& Owocki~\cite{cranmer} and Owocki \& Gayley\cite{owocki4} studied the effects of limb darkening, gravity darkening and oblateness on the radiation driving mechanism and found that  the tendency for the higher flux from the bright poles to drive material toward the darker equatorial region is outweighed by the opposite tendency for the oblateness of the stellar surface to direct the radiative flux to higher latitudes, i.e. away from the equator. The paper review by Porter \& Rivinius~\cite{porter} also outline the effect of the inclusion of nonradial line-driving force which reduces the effect of the wind compression to zero and, taking into account the gravity darkening, results in a polar wind enhancement.  This physical effect goes in the opposite direction to the one proposed for the Wind Compressed Disk model from Bjorman \& Cassinelli~\cite{Bjorkman93}. In a recent paper, Kervella \& Domiciano de Souza~\cite{kervella} have evidenced an enhanced polar wind for the Be star Achernar whereas this Be star presents no 
hydrogen lines in strong emission. Thus, it seems that a significant polar wind may be present even if the star is still in a normal B phase, i.e. this enhanced polar wind does not seem to be related to the existence of a dense equatorial envelope, as already outlined by Kervella \& Domiciano de Souza~\cite{kervella}.

\section{Conclusion}
\label{conclusion}

\begin{enumerate}
      \item Thanks to these first spectrally resolved interferometric measurements of a Be star at 2 $\mu$m we are able to propose a possible scenario for the Be star $\alpha$ Arae's circumstellar environment which consist in a thin disk + polar enhanced winds that is successfully modeled with the SIMECA code.      
       \item We found that the disk around $\alpha$ Arae is compatible with a dense equatorial  matter confined in the central region whereas a polar wind is contributing along the rotational axis of the central star. Between these two regions the density must be low enough to reproduce the large visibility modulus (small extension) obtained for two of the four VLTI baselines. This new scenario is also compatible with the previous MIDI measurements and the fact that the outer part of the disk may be truncated by a unseen companion at 32~R$_\star$.   
      \item We obtain for the first time the clear evidence that the disk is in Keplerian rotation, closing a debate that occurs since the discovery of the first Be star $\gamma$ Cas by father Secchi in 1866.
       \item We found that that $\alpha$ Arae must be rotating very close to its critical velocity.
        \item These observations were done using the medium (1500) spectral resolution of the VLTI/AMBER instrument and are very promising for the forthcoming AMBER high spectral resolution observational mode (10000) and the coupling of the Auxiliary Telescopes (ATs) on the VLTI array.                        
\end{enumerate}


\begin{acknowledgements}
This research has made use of SIMBAD database, operated at CDS, Strasbourg, France. We thanks G. Duvert and J.M. Clausse for their help in the data reduction through the JMMC. We acknowledge the useful comments of the referee Thomas Rivinius who really help to improve the discussion in this paper.
\end{acknowledgements}

\end{document}